\documentclass[prd,preprint,eqsecnum,nofootinbib,amsmath,amssymb,
               tightenlines,dvips,superscriptaddress,floatfix]{revtex4}
\usepackage{graphicx}
\usepackage{bm}
\usepackage{amsfonts}
\usepackage{amssymb}

\usepackage{dsfont}

\def\Nc{N_{\rm c}}
\def\Nf{N_{\rm f}}
\def\CA{C_{\rm A}}

\def\alphas{\alpha_{\rm s}}
\def\alphaqed{\alpha_{\scriptscriptstyle\rm EM}}
\def\Re{\operatorname{Re}}

\def\ix{{\rm i}}

\def\yfrake{{\mathfrak y}_e}

\def\gammaE{\gamma_{\rm\scriptscriptstyle E}}

\def\xe{x_e}
\def\ye{y_e}

\def\MSbar{\overline{\mbox{MS}}}

\def\testnum{\chi}
\def\test{\testnum\alpha}
\def\lstop{\ell_{\rm stop}}
\def\tildelstop{{\tilde\ell}_{\rm stop}}
\def\alphaMSbar{\alpha_{\overline{\scriptscriptstyle{\rm MS}}}}
\def\Avg{\operatorname{Avg}}

\begin {document}



\title
    {
      Strong- vs.\ weak-coupling pictures of jet quenching:\\
      a dry run using QED
    }

\author{Peter Arnold}
\affiliation
    {%
    Department of Physics,
    University of Virginia,
    Charlottesville, Virginia 22904-4714, U.S.A.
    \medskip
    }%
\author{Shahin Iqbal}
\affiliation
    {%
    National Centre for Physics,
    Quaid-i-Azam University Campus, \\
    Shahdra Valley Road,
    P.O. Box No. 2141,
    Islamabad -- 440000, Pakistan
    \medskip
    }%
\author{Tanner Rase}
\affiliation
    {%
    Department of Physics,
    University of Washington,
    Seattle, Washington 98195, U.S.A.
    \medskip
    }%

\date {\today}

\begin {abstract}%
{%
  High-energy partons ($E \gg T$) traveling through a quark-gluon plasma
  lose energy by splitting via bremsstrahlung and pair production.
  Regardless of whether or not the quark-gluon plasma itself is
  strongly coupled, an important question lying at the heart
  of philosophically different approaches to energy loss is whether
  the high-energy partons of an in-medium shower can be thought of
  as a collection of individual particles, or whether
  their coupling to each other is also so strong that a
  description as high-energy ``particles'' is inappropriate.
  We discuss some possible theorists' tests of this question
  for simple situations (e.g.\ an infinite, non-expanding plasma)
  using thought experiments and first-principles
  quantum field theory calculations
  (with some simplifying approximations).
  The physics of in-medium showers is substantially affected by
  the Landau-Pomeranchuk-Midgal (LPM) effect, and our proposed
  tests require use of what might be called
  ``next-to-leading order'' LPM results, which account for
  quantum interference between consecutive splittings.
  The complete set of such results is not yet available for
  QCD but is already available for the theory of large-$\Nf$
  QED.  We therefore use large-$\Nf$ QED as an example, presenting
  numerical results
  as a function of $\Nf\alpha$, where
  $\alpha$ is the strength of the coupling at the relevant high-energy
  scale characterizing splittings of the high-energy particles.
}%
\end {abstract}

\maketitle
\thispagestyle {empty}

{\def\boldmath{}\tableofcontents}
\newpage


\section{Introduction and Preview of Results}
\label{sec:intro}

\subsection{Overview}

\begin {figure}[ht]
\begin {center}
  \includegraphics[scale=0.3]{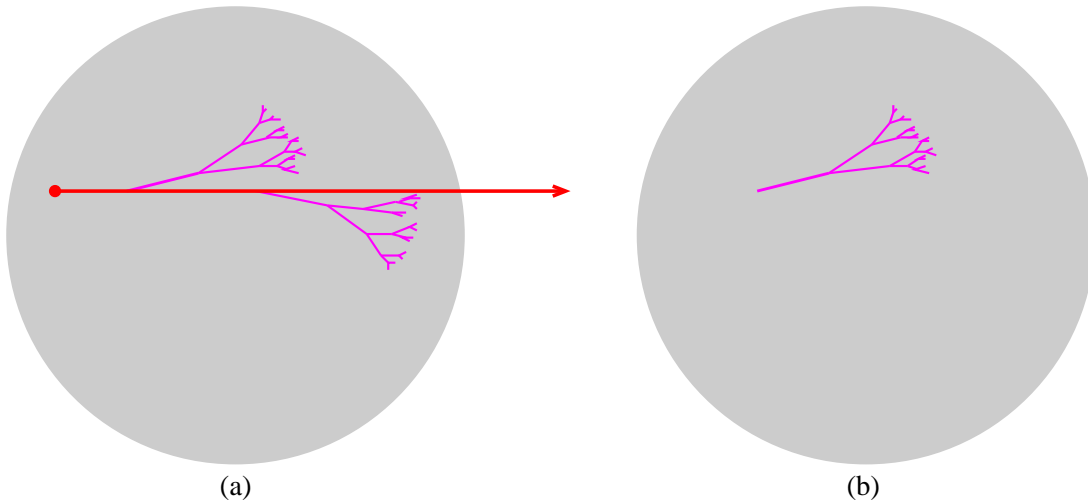}
  \caption{
     \label{fig:cartoon}
     (a) A cartoon of in-medium energy loss of a high-energy parton
     traversing a quark-gluon plasma.  (Initial vacuum-like radiation
     associated with the hard event that created the initial particle
     is not shown.  Vacuum-like radiation and then hadronization of
     a high-energy particle that leaves the medium is also not shown.) 
     (b) Depiction of a branch that deposits all of its energy in
     the medium.
  }
\end {center}
\end {figure}

Consider the cartoon in fig.\ \ref{fig:cartoon}a depicting energy loss
of a high-energy parton traversing a lower-energy quark-gluon plasma
(QGP) produced in a relativistic heavy-ion collision.  This
figure and this paper focuses on the in-medium evolution,
ignoring the complications of both (i)
early-time initial vacuum-like radiation associated with the
hard parton-level collision that scattered the high-energy parton out of
the beam direction in the first place and (ii) the late-time
hadronization when a high-energy parton leaves the quark-gluon plasma.
Generically, energy loss of high-energy relativistic particles
traveling through a medium is dominated by bremsstrahlung and pair
production, as depicted by the splittings in the cartoon.  In detail,
this picture of energy loss implicitly assumes that the high-energy
particle and its high-energy daughters and granddaughters {\it et cetera}
may be thought of as
individual particles, which in quantum field theory implies a perturbative
description of the {\it states} of the high-energy particles
(even if the underlying quark-gluon plasma
is strongly coupled).
Such a perturbative treatment requires that the running QCD
coupling $\alphas(\mu)$ be ``small'' at the relevant scale $\mu$ that
characterizes the high-energy splittings.
This is in contrast to 
the separate question of whether the quark-gluon plasma itself is
weakly- or strongly-coupled, which depends on the strength of
$\alphas$ at QGP scales, such as $\alphas(T)$ or $\alphas(\xi^{-1})$,
where $T$ is the temperature and $\xi$ is the QGP chromo-electric
screening length.

Though real-world quark-gluon plasmas are generally considered to be strongly
coupled (or to at the very least require complicated
reorganizations of small-coupling expansions \cite{wQGP}),
the situation has been less clear with
regard to the coupling of each high-energy parton to its daughters in
the energy-loss picture of fig.\ \ref{fig:cartoon}.  On the
``weakly-coupled splitting'' side, there is a large literature built
upon the formalism of Baier et al.\ \cite{BDMPS12,BDMPS3} and
Zakharov \cite{Zakharov} (BDMPS-Z), which
treats high-energy partons as individual particles.
At high energy, the dependence of splitting on medium properties may
(with caveats) for many purposes
be summarized by the medium parameter $\hat q$, whether the medium itself
is weakly or strongly coupled.
$\hat q$ is the proportionality constant in the relationship
$\langle Q_\perp^2 \rangle = \hat q \, \Delta z$, where
$Q_\perp$ is the transverse momentum that a high-energy
parton picks up, relative to its initial direction of motion,
after passing through length $\Delta z$ of the medium.
Alternatively, on the
``strongly-coupled splitting'' side, where high-energy partons cannot
be treated individually, there are studies of how energy loss behaves
at extremely large coupling in QCD-like theories with gravity duals
\cite{GubserGluon,HIM,CheslerQuark}.%
\footnote{
  For an example of a phenomenological model that attempts to combine
  certain aspects of both the weakly-coupled and strongly-coupled pictures
  of splitting, see the hybrid model of ref.\ \cite{hybrid}.
}
Because of all the complexities that enter describing the
many different stages of relativistic heavy-ion collisions, it has
been difficult to settle the issue of weakly vs.\ strongly coupled
splitting from experimental data.
So it would
be useful to have theoretical calculations we could do,
for some simplified thought experiments, that would shed light on
whether we are in the weakly-coupled or strongly-coupled splitting
limit for energy loss.

As we shall review, previous authors \cite{Blaizot,Iancu,Wu}
have made leading-log studies
of the correction to BDMPS-Z splitting rates due to additional,
softer gluon bremsstrahlung happening simultaneously with the underlying
splitting process.  They found that such corrections are large
but are also universal in the sense that they are completely absorbed
(at leading log level) by accounting for similar corrections to
$\hat q$.  One may then pose a refined question about the merits
of a weakly-coupled splitting picture of in-medium showering:
how large are the corrections to this picture that {\it cannot} be
absorbed into an effective value of $\hat q$?%
\footnote{
  There is precedent for a similar distinction in a
  very different context:
  For calculations of a weakly-self-interacting
  quark gluon plasma [that is, $\alphas(gT)$ small, which we do not
  assume here],
  calculations of the shear viscosity
  to entropy ratio $\eta/s$ of the quark-gluon plasma find that
  the next-to-leading order (NLO) correction in $\alphas(gT)$
  is very large but almost completely accounted
  for by the also-large NLO correction to $\hat q$ \cite{NLOshear}.
  (Technically, the calculation covers only
  ``almost'' all next-to-leading order contributions to $\eta/s$
  \cite{NLOshear}.)
  We should emphasize that their expansion in the quark-gluon plasma
  coupling is not our expansion in the high-energy splitting coupling, and
  what they call a NLO correction to $\hat q$ is something that in our
  expansions will already be part of the leading-order value of $\hat q$,
  which we will later refer to as $\hat q^{(0)}$.
}

In this paper, we will give some examples of measures that can be
used in theory thought experiments to address this question, and
we will discuss advantages and disadvantages of those measures.
Though our ultimate motivation is to eventually address the question for QCD,
the full set of QCD calculations required is not yet available.
Here we will show concrete results for a different theory: large-$\Nf$ QED.
For that case, we give quantitative
measures of the size of corrections to the weakly-coupled splitting
picture as a function of the size of $\Nf \alphaqed$.

For the sake of simplicity, we are going to
focus on high-energy particles which completely stop and deposit
their energy in the medium, like the first branch of
fig.\ \ref{fig:cartoon}a, isolated in fig.\ \ref{fig:cartoon}b.
Phenomenologically, one could view this as a choice to focus
on partons whose energy is large compared to that of typical
plasma particles but not so large that they punch all the way
through the plasma.  Alternatively, one could instead consider
arbitrarily high energy partons but imagine, as a theory exercise,
that the extent
of the plasma were large enough for them to stop completely.
Whether in-medium showers of high-energy particles
have weakly-coupled or strongly-coupled splittings in
this case is an interesting and important
first question.  And studying that limit provides a way to
examine the issue independent of the complicated details necessary
to discuss all cases relevant to the full phenomenology of
energy loss.

Throughout, we will use the phrase ``quark-gluon plasma'' as a
generic term for the QCD background generically produced in
heavy-ion collisions without committing to any more specific, detailed
picture of that background other than that scattering of high-energy
particles from the medium can be characterized by some value of
$\hat q$.


\subsection{Overlapping Formation Times}

\begin {figure}[t]
\begin {center}
  \includegraphics[scale=0.5]{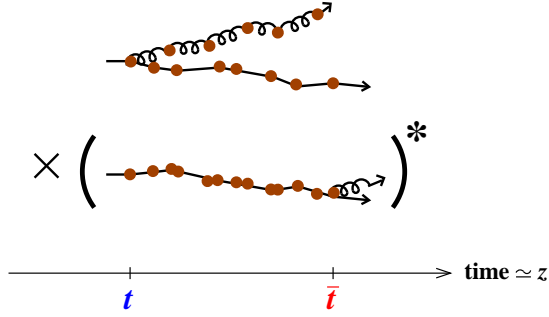}
  \caption{
     \label{fig:lpmsplit}
     Interference between high-energy bremsstrahlung at time $t$ in
     the amplitude and time $\bar t$ in the conjugate amplitude,
     as the high-energy particles repeatedly scatter
     (represented by the brown dots) by small angles from the medium.
     In QCD, the particle represented by the non-curly line could be a quark
     or a gluon. For photon bremsstrahlung, photon scattering from the
     medium can be ignored.
  }
\end {center}
\end {figure}

In any single example of a shower, the exact time or place where each splitting
occurred has some ambiguity, and the extent of that uncertainty
is known as the formation time or formation length for that splitting.
In terms of calculations of the splitting rate, consider multiplying
an amplitude times a conjugate amplitude to get a rate.
The formation time corresponds, parametrically,
to the largest time difference $|t-\bar t|$ for which a splitting at time
$t$ in the amplitude non-negligibly interferes with the same splitting
at time $\bar t$ in the conjugate amplitude, as depicted in
fig.\ \ref{fig:lpmsplit}.  At high energy,
the corresponding formation length in the direction of motion
is the same as the formation time
(times the speed of light), and they grow with
energy.  For example, for bremsstrahlung with energy $xE$ from
a parton with energy $E$, the formation length is parametrically%
\footnote{
   For a very brief review of formation times in the
   language of this paper, see section II.A.1 of ref.\ \cite{QEDnf}.
   The QED formation time essentially goes back to Migdal \cite{Migdal},
   though he did not use this language.  ($\hat q t_{\rm form}$ is
   proportional to what Migdal would call $m^2/s$ for bremsstrahlung
   and $m^2/\bar s$ for pair production.)
}
\begin {subequations}
\label {eq:lform}
\begin {align}
   l_{\rm form} &\sim \sqrt{ \frac{x(1-x)E}{\hat q} }
   \qquad \mbox{(gluon bremsstrahlung)} ,
\label {eq:lformG}
\\
   l_{\rm form} &\sim \sqrt{ \frac{(1-x)E}{x\hat q} }
   \qquad \mbox{(photon bremsstrahlung)} ,
\end {align}
\end {subequations}
in the limit of a thick medium.
When the formation time becomes large compared to the mean free
path for collisions with the medium, as in fig.\ \ref{fig:lpmsplit},
it causes a reduction in the bremsstrahlung rate known as the
Landau-Pomeranchuk-Migdal (LPM) effect \cite{LP,Migdal}.
A similar reduction occurs
for pair production.

\begin {figure}[t]
\begin {center}
  \includegraphics[scale=0.6]{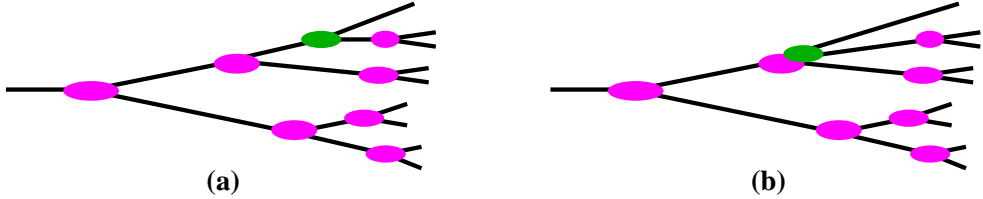}
  \caption{
     \label{fig:shower}
     An example of an in-medium shower like fig.\ \ref{fig:cartoon}b,
     but here depicting the ambiguity of the longitudinal position of
     each splitting by an oval that represents the formation length.
     (a) shows the case where there is no interference between splittings;
     (b) shows a case where two consecutive splittings overlap.
  }
\end {center}
\end {figure}

In fig.\ \ref{fig:shower}a, we show a shower where we have
qualitatively depicted
the formation lengths for each splitting by ovals.  This picture
implicitly assumes a hierarchy of scales in which the
distance between consecutive splittings is large compared to
the formation lengths for each splitting, so that the probability of
each splitting is independent.  If true,
one may take as a starting point for simulating this type of
in-medium shower a calculation (or model) of single-splitting rates.
As simulation time progresses, one could roll dice every
time increment to determine whether a new splitting happened in
that time interval.  Such a model of independent dice rolls for each
splitting assumes that the chance of two or more consecutive splittings
having {\it overlapping formation} times, such as depicted in
fig.\ \ref{fig:shower}b, is small.


\begin {figure}[t]
\begin {center}
  \begin{picture}(300,80)(0,0)
  \put(0,15){\includegraphics[scale=0.5]{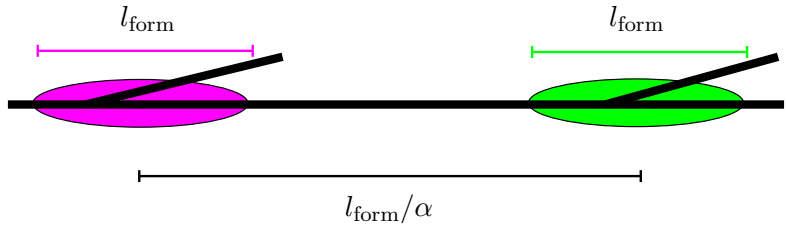}}
  \put(130,3){$l_{\rm form}/\alpha$}
  \put(45,75){$l_{\rm form}$}
  \put(230,75){$l_{\rm form}$}
  \end{picture}
  \caption{
     \label{fig:hierarchy}
     The hierarchy of scales (\ref{eq:lrad}).
  }
\end {center}
\end {figure}

Consider for a moment the case of nearly-democratic splittings,
where the two daughters of each splitting have roughly comparable
energy.  In general, at very high energy,
each formation length of scattering from the medium offers one chance for
splitting, and the chance of such a splitting is parametrically
of order the coupling $\alpha$ associated with the splitting
vertex.  The typical distance $l_{\rm rad}$ between consecutive splittings
is then
\begin {equation}
   l_{\rm rad} \sim \frac{l_{\rm form}}{\alpha} \,.
\label {eq:lrad}
\end {equation}
If the relevant value of $\alpha$ is small enough, this leads to
the hierarchy of scales shown in fig.\ \ref{fig:hierarchy}, leading
to the picture of independent splittings in fig.\ \ref{fig:shower}a.
The probability of any two particular splittings overlapping, like
the two in fig.\ \ref{fig:shower}b, would be $O(\alpha)$.

Since $\alpha$ runs and depends on scale,
we need to identify the relevant renormalization
scale for the $\alpha$ in (\ref{eq:lrad}).
That $\alpha$ characterizes the strength
of the vertex where the high-energy parent splits into its high-energy
daughters, and the scale we need is therefore
the distance scale characterizing
the separation of the daughters during the splitting process,
i.e.\ during the formation time.
That separation $b$ is transverse and so is related by
the uncertainty principle $Q_\perp \sim 1/b$ to the
relative transverse momentum $Q_\perp \sim \sqrt{\hat q t_{\rm form}}$
that the high-energy particles
pick up during the splitting process.
So we are interested in $\alpha(Q_\perp)$,
with (\ref{eq:lform}) giving
\begin {equation}
  Q_\perp \sim (\hat q E)^{1/4} .
\label {eq:Qperp}
\end {equation}
The above argument that (\ref{eq:Qperp}) is the relevant scale
for $\alpha$ is qualitative, but
it has been checked by explicit calculation in the case of
large-$\Nf$ QED \cite{QEDnf}.

Note that though this scale grows with energy $E$, it does not grow
quickly!  For example, for $\hat q \sim 1$ GeV$^3$ for quark-gluon
plasmas and $E$ as large as $\sim 100$ GeV,
the parametric estimate (\ref{eq:Qperp}) is $Q_\perp \sim$ only a few GeV
(in the theoretical limit of a thick medium).
If $\alpha(Q_\perp)$ is too big, then splittings will usually overlap,
and we cannot then use a weakly-coupled description of splittings.
A way to diagnose this problem
is to first formally assume $\alpha$ is small, calculate
the corrections to shower development due to overlapping formation
times, and then see how large those corrections are when one evaluates
them for realistic values of $\alpha$.  If the corrections are of order
20\%, for example,
then the basic picture of weak-splitting behind figs.\ \ref{fig:cartoon}
and \ref{fig:shower}a is reasonable.  If the corrections are 200\%,
then it is not, and the number of high-energy particles present
at any moment in the shower would not (even approximately) be a
well-defined concept.

Throughout this paper, the coupling $\alpha$ will generally refer to
$\alpha(Q_\perp)$ [and not to $\alpha(T)$].


\subsection{Stopping distances}

Hard bremsstrahlung is more efficient at decreasing the energy of
particles in the shower than soft bremsstrahlung.  In more
general language, nearly-democratic splittings are more
efficient for development of the shower than splittings where
one of the daughters is soft.  To study the size of
overlap effects on the development and stopping of showers,
we would like to find simple quantities that naturally weight different
processes by how much they contribute to the degradation of the
energies of the particles in the shower.  A natural quantity to consider
is simply the characteristic length of the shower to where it
ends and deposits its energy into the medium.  We'll generically
refer to this as the average ``stopping distance'' $\ell_{\rm stop}$.

There are different variants of stopping length one could define.
If the original high-energy particle carries a conserved charge,
like fermion number, then one could track the statistical distribution
$\rho(z)$ of the charge density deposited in the
medium as a function of distance $z$ from the initial position
of the initial high-energy particle.%
\footnote{
  Once deposited, charge will then flow with the medium and also
  diffuse hydrodynamically.  Our definition here of $\rho(z)$ does
  not follow that evolution
  but instead refers to where the charge
  was deposited when it became part of the medium.
}
(Transverse displacements will
be small compared to $z$, so we need not distinguish between
distance and longitudinal distance.)
Following ref.\ \cite{stop}, we then define the
average ``charge stopping distance''
as the first moment
of the distribution:
\begin {subequations}
\label {eq:lstopdef}
\begin {equation}
   \ell_{\rm stop}^{\rm (charge)}
   \equiv \langle z \rangle_{\rho}
   \equiv
   \frac{1}{q_0} \int_0^\infty dz \> z \, \rho(z)
   ,
\end {equation}
where $q_0$ is the charge of the initial particle (e.g.\ $q_0{=}1$).
Alternatively, we could do the same with the energy density
$\varepsilon(z)$ deposited in the medium to define an
``energy stopping distance,''
\begin {equation}
   \ell_{\rm stop}^{\rm (energy)}
   \equiv \langle z \rangle_{\varepsilon}
   \equiv
   \frac{1}{E} \int_0^\infty dz \> z \, \varepsilon(z)
   .
\end {equation}
\end {subequations}

In principle, the very last stage for particles from the shower
stopping in the medium---when their energies finally become comparable to
the characteristic energy scale $T$ of the plasma---is
not described by the high-energy
approximations that we shall use throughout this paper.  Fortunately,
the effect of that last stage has only a parametrically
small effect on the stopping length for showers initiated by
high-energy particles $E \gg T$
\cite{stop}.  To see this, consider the scale
(\ref{eq:lrad}) characterizing the distance between consecutive
splittings.  For nearly-democratic splittings, this is
\begin {equation}
   l_{\rm rad}(E) \sim 
  \frac{1}{\alpha}
  \sqrt{ \frac{E}{\hat q} } .
\label {eq:lrad1}
\end {equation}
As a crude estimate, imagine that the energies of individual particles
drop by a factor of 2 with each splitting.  Then the total distance
the shower would propagate would be
\begin {align}
   \ell_{\rm stop}(E) &\sim 
   l_{\rm rad}(E)
   + l_{\rm rad}(\tfrac12 E)
   + l_{\rm rad}(\tfrac14 E)
   + l_{\rm rad}(\tfrac18 E)
   + \cdots
\nonumber\\
   &\sim
   \frac{1}{\alpha} \sqrt{ \frac{E}{\hat q} }
   +
   \frac{1}{\alpha} \sqrt{ \frac{E}{2\hat q} }
   +
   \frac{1}{\alpha} \sqrt{ \frac{E}{4\hat q} }
   +
   \frac{1}{\alpha} \sqrt{ \frac{E}{8\hat q} }
   + \cdots .
\label {eq:series}
\end {align}
This is a convergent series giving, parametrically,
\begin {equation}
   \ell_{\rm stop}(E)
   \sim \frac{1}{\alpha} \sqrt{\frac{E}{\hat q}}
   \sim l_{\rm rad}(E) .
\label {eq:lstopparam}
\end {equation}
The error in the treatment of stopping when the particles
degrade to energy of order $T$ is just a piece of order
$l_{\rm rad}(T) \sim \alpha^{-1} \sqrt{T/\hat q}$ in the series (\ref{eq:series})
and so is suppressed by $\sqrt{T/E}$ compared to the sum
(\ref{eq:lstopparam}).

We could now ask what effect overlapping formation times have on these stopping
lengths.  Is it a small or large effect for relevant sizes of the
coupling $\alpha$?  Consider an expansion
\begin {equation}
   \ell_{\rm stop} =
   \ell_{\rm stop}^{(0)} + \Delta \ell_{\rm stop}
   + O(\alpha^2 \ell_{\rm stop}) .
\label {eq:expansion1}
\end {equation}
Here $\ell_{\rm stop}^{(0)}$ is the result in the approximation that
no splittings interfere with each other,
in which case the statistical development
of the shower can be computed based just on single-splitting probabilities.
$\Delta \ell_{\rm stop}$ is the $O(\alpha \ell_{\rm stop})$ correction
due to overlapping
formation lengths, computed in the formal limit
that $\alpha$ is small.  We might then consider taking the ratio
\begin {equation}
   \frac{\Delta \ell_{\rm stop}}{\ell_{\rm stop}^{(0)}} = O(\alpha)
\label {eq:ratio0}
\end {equation}
as our measure of whether splitting in showers is weakly or
strongly coupled for a given size of $\alpha$.

\begin {figure}[t]
\begin {center}
  \includegraphics[scale=0.5]{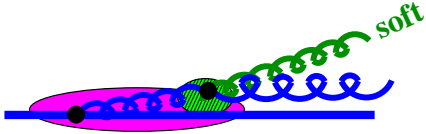}
  \caption{
     \label{fig:soft}
     An example of overlapping formation lengths associated with
     soft gluon emission from a harder bremsstrahlung
     gluon.  For QCD, softer gluon emission corresponds to shorter
     formation lengths (the green region) than harder gluon emission
     (the magenta region).  This type of process contributes to the double
     logarithm in (\ref{eq:dbllog0}).
  }
\end {center}
\end {figure}

There is a problem with this measure for QCD.
In the parametric estimates
leading to (\ref{eq:ratio0}),
we have so far only discussed nearly-democratic splitting.  However, soft
gluon bremsstrahlung ($x{\ll}1$)
is associated with shorter formation lengths than
hard gluon bremsstrahlung, as reflected in the small $x$ dependence
of (\ref{eq:lformG}).  That means that soft gluon bremsstrahlung is
less LPM suppressed and so happens more often than hard gluon
bremsstrahlung.  Various authors \cite{Blaizot,Iancu,Wu} have found that overlap
effects of soft-gluon bremsstrahlung during harder-gluon bremsstrahlung,
as depicted in
fig.\ \ref{fig:soft}, correct the harder emission probability by
an amount that is formally suppressed by order%
\footnote{
  If the medium is expanding,
  ref.\ \cite{ExpandingDblLog} shows in detail that
  the $L$ in (\ref{eq:dbllog0L}) should be replaced by
  $\sim \min(L,\tau)$ where $\tau$ is the characteristic time of
  expansion (at the time of the splitting).
  In our paper, we consider a static medium for
  simplicity of designing theory tests of weak vs.\ strong splitting.
}
\begin {equation}
  \alphas \ln^2\Bigl(\frac{L}{\tau_0}\Bigr)
\label {eq:dbllog0L}
\end {equation}
instead of $\alphas$, where $\tau_0$
characterizes the scale of the mean-free path for scattering
from the medium and reflects a characteristic scale of the medium.
(Here we write $\alphas$ instead of $\alpha$ because this part of
the discussion is specific to QCD and does not apply to QED.)
In the context of a thick medium, as considered in this paper,
the $L$ in the logarithm should be interpreted as the formation
time for the harder splitting.
If the harder splitting is nearly
democratic, then $L \sim \sqrt{E/\hat q}$ and so the effect
of a soft emission during a nearly-democratic splitting is
suppressed by order
\begin {equation}
   \alphas \ln^2 \Bigl( \frac{E}{\hat q \tau_0^2} \Bigr) ,
\label {eq:dbllog0}
\end {equation}
which is no suppression at all for energy large enough that
the double log compensates for the smallness of $\alphas(Q_\perp)$.

This effect on splitting rates will result in
the same soft double-log enhancement to the
ratio considered in (\ref{eq:ratio0}), so that
\begin {equation}
   \frac{\Delta \ell_{\rm stop}}{\ell_{\rm stop}^{(0)}}
   = O\bigl(\alphas \ln^2(E/\hat q\tau_0)\bigr) .
\label {eq:lstopRatio0}
\end {equation}
However, the double-log corrections to splitting rates have a very special
form \cite{Blaizot,Iancu,Wu}:
they can be absorbed into a redefinition of the effective value
of $\hat q$, exactly corresponding to the result of an earlier
calculation \cite{Wu0} of the double-log
soft-bremsstrahlung correction $\delta \hat q$
to the definition of
$\hat q$ itself.  At first order in $\alpha$, in leading-log
approximation, they found
\begin {equation}
  \delta \hat q_R \approx
  \frac{\CA\alphas}{2\pi} \, \hat q_R^{(0)} \ln^2\Bigl(\frac{L}{\tau_0}\Bigr)
  ,
\label {eq:deltaqhat}
\end {equation}
where $R$ indicates the color representation of the particle whose
deflection is being described by $\hat q$,
where $\CA{=}\Nc$ is the quadratic Casimir for the adjoint color
representation (i.e.\ for a bremsstrahlung gluon), and where
$\approx$ is our notation for leading-log approximation.
Moreover, the leading logs can be summed to all orders in $\alphas$ \cite{Wu0}
with the result that the effective value of $\hat q$ scales with $L$ as
$\hat q_{\rm eff} \sim L^\gamma$ \cite{Blaizot} for sufficiently large
values of $L$, with
$\gamma = 2 \sqrt{\CA\alphas/\pi}$.  For splitting in
the limit of a thick medium, this corresponds to the scaling
\begin {equation}
   \hat q_{\rm eff}(E) \sim E^{\kern1pt\gamma/2} = E^{\sqrt{\CA\alphas/\pi}}
\end {equation}
for the effective $\hat q$ to use in
calculations of nearly-democratic splitting.
If inserted into
the parametric formula (\ref{eq:lstopparam}) for the stopping length,
this gives
\begin {equation}
   \ell_{\rm stop} \propto E^{\frac12 (1-\frac{\gamma}{2})}
   \sim E^{\frac12 \bigl[1 - \sqrt{\CA\alphas/\pi}\,\bigr]} ,
\label {eq:lstop1}
\end {equation}
up to yet higher order corrections in the exponent.
This is an important result for theory because it explains how,
in QCD-like theories where strong-coupling results are also known
(namely theories that have gravity duals),
the weak-coupling behavior $\ell_{\rm stop} \propto E^{1/2}$ can
change as $\alphas$ is increased to move toward the known strong-coupling
behavior $\ell_{\rm stop} \propto E^{1/3}$ \cite{GubserGluon,HIM,CheslerQuark}.
The above $E^{-\gamma/4}$ relative correction from soft overlapping emissions
to the weak-splitting result $E^{1/2}$ is a parametrically large effect
at large $E$, even if $\gamma$ were small.
Moreover, the relative correction $\gamma/2$ to the exponent is controlled by
the size of $\sqrt{\alphas}$ rather than (the parametrically smaller)
$\alphas$.

The upshot is that it has already been known for some time that
corrections due to overlapping formation lengths are very significant
in the case of overlap with soft emissions, but that this effect
can be absorbed into an effective value of $\hat q$.
On the more phenomenological side, Zakharov \cite{Zakharovqhat}
has recently calculated
that other, {\it non}-logarithmic corrections to $\hat q$
at relative order $\alphas$
are comparably important in applications of interest and can change
the sign of the effect.

In this paper, we want to address the question of how large are the
corrections due to overlapping formation times that {\it cannot}
simply be absorbed into an effective value of $\hat q$?  There are two
ways to do this.  One is to look for some different measure that is
insensitive to the size of $\hat q$ in the first place and so avoids
double-log enhanced corrections.  We'll propose a (partly) successful
choice in section \ref{sec:sigma} below.

Alternatively, one can re-organize the expansion (\ref{eq:expansion1})
so that the $\hat q$ used in the leading term is calculated using the
effective value $\hat q_{\rm eff}$ of $\hat q$ instead of the original
value $\hat q^{(0)}$, where
$\hat q_{\rm eff} = \hat q^{(0)} + \delta \hat q$
when formally expanded to first order in $\alpha$.  There is
an ambiguity in exactly how one defines $\hat q_{\rm eff}$ in this
context, however, because the identification of the scale $L$ in
(\ref{eq:deltaqhat}) as the formation length $\sim \sqrt{E/\hat q}$ is
only a parametric identification of scale.  Similar to ambiguities
associated with choosing renormalization scale in perturbation theory,
or factorization scales for defining parton distribution functions at
next-to-leading order, the exact choice of $L$ here is a matter of
convention as long as one chooses $L \sim \sqrt{E/\hat q}$
in order to eliminate the large double logarithm in the re-organized
small-$\alpha$ expansion.  (There are further concerns one can have
about sub-leading, single log terms, but we leave
that for later discussion.)

\begin {figure}[t]
\begin {center}
  \includegraphics[scale=1.0]{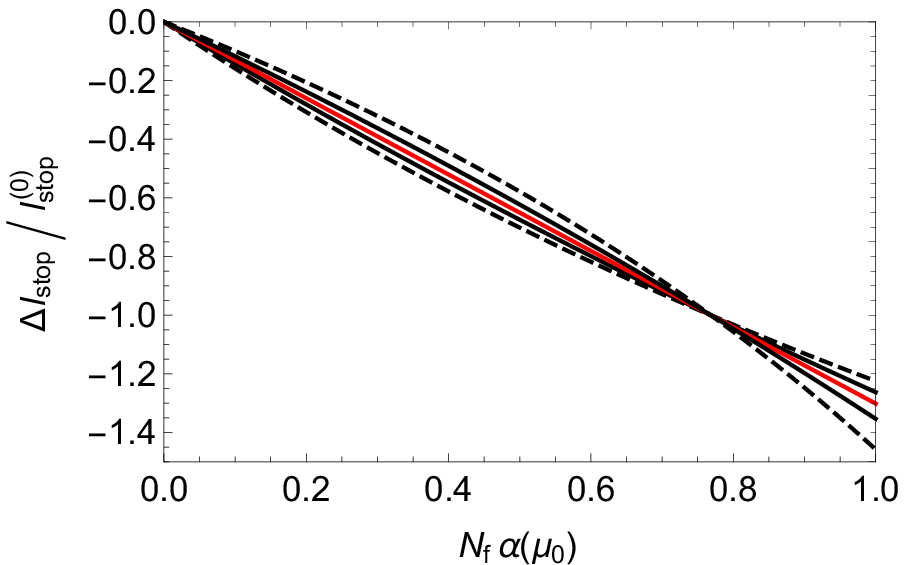}
  \caption{
     \label{fig:lstopPlot}
     The relative size $\Delta\lstop/\lstop^{(0)}$ of the NLO correction
     in large-$\Nf$ QED to
     charge stopping length $\lstop^{\rm charge}$,
     versus $\Nf \alphaMSbar(\mu_0)$ for $\mu_0 \equiv (\hat q E)^{1/4}$.
     The different lines correspond to different choices
     ($\mu_0/5$, $\mu_0/2$, $\mu_0$, $2\mu_0$, $5\mu_0$) of
     the renormalization scale $\mu$ used in the calculation of
     $\Delta\lstop/\lstop^{(0)}$.
     The features of this graph are discussed in
     section \ref{sec:RenormScale}.
  }
\end {center}
\end {figure}

Though we have motivated our discussion with QCD applications, in this
paper we are going to give example calculations for the case of
large-$\Nf$ QED, where $\Nf$ is the number of electron flavors.  For
this theory, all the overlap corrections to BDMPS-Z splitting rates
that will be needed are already available \cite{QEDnf}.  (The only
reason for the large-$\Nf$ limit was to reduce the number of diagrams
that needed to be calculated in ref.\ \cite{QEDnf}.)
The behavior of the LPM effect in QED is qualitatively similar to
the LPM effect in QCD for the case of nearly-democratic bremsstrahlung
or pair production but is qualitatively different for the case of
softer bremsstrahlung.  In QCD there is less LPM suppression of softer
bremsstrahlung; in QED there is more.  In particular, QED does not have
the double-log enhancement (\ref{eq:lstopRatio0}).
It does, however, have a different type of logarithm (which QCD
also has):
for any quantity that depends on coupling $\alpha$ at leading order,
the next-to-leading-order (NLO) correction will necessarily have
logarithmic dependence on the choice of
renormalization scale.
As we'll discuss, this logarithm has the form
\begin {equation}
   \frac{\Delta \ell_{\rm stop}}{\ell_{\rm stop}^{(0)}}
   = O\Bigl(\Nf\alphaqed \ln\bigl(\tfrac{\mu}{(\hat q E)^{1/4}}\bigr)\Bigr)
      + O(\Nf\alphaqed) .
\label {eq:lstopRatio0qed}
\end {equation}
As discussed qualitatively before in the context of (\ref{eq:Qperp}), the
renormalization scale should be chosen of order
$Q_\perp \sim (\hat q E)^{1/4}$, here in order to eliminate large
logarithms (\ref{eq:lstopRatio0qed})
in the expansion of $\ell_{\rm stop}$ in $\alpha$.
But, like our discussion of $L$ for absorbing QCD double logs,
the exact choice to make
for $\mu$ is ambiguous.  Since our goal is to judge the relative
size of NLO corrections, we must therefore account for this ambiguity
by considering a reasonable range of $\mu$.
Fig.\ \ref{fig:lstopPlot} previews our large-$\Nf$ QED results for
the size of the relative correction $\Delta\lstop/\lstop$
to the charge stopping length $\ell_{\rm stop}^{\rm charge}$
as a function of $\Nf\alphaqed$,
where the different curves give a sense of the ambiguities associated with
varying the choice of $\mu$ around $(\hat q E)^{1/4}$.
The central line is given by
\begin {equation}
   \Delta\lstop/\lstop^{(0)} =
   -1.302 \, \Nf\alphaqed\bigl((\hat q E)^{1/4}\bigr) .
\end {equation}
We will comment at the end of the paper on what one might take away
from this and other QED results.

\subsection{Shape of stopping distribution}
\label {sec:sigma}

Ideally, it would be nice if in the QCD case we could sidestep the
scale ambiguities associated with absorbing the corrections to $\hat q$ and
instead construct a thought-experiment observable that is independent
of the size of $\hat q$.
We have a (partly) successful proposal
(we'll explain the caveat ``partly'' later on)
which has the added benefit of pushing the renormalization-scale
ambiguity of $\alpha(\mu)$ off to next-to-next-to-leading order
(NNLO).

\begin {figure}[t]
\begin {center}
  \begin{picture}(290,155)(0,0)
  \put(17,0){\includegraphics[scale=1]{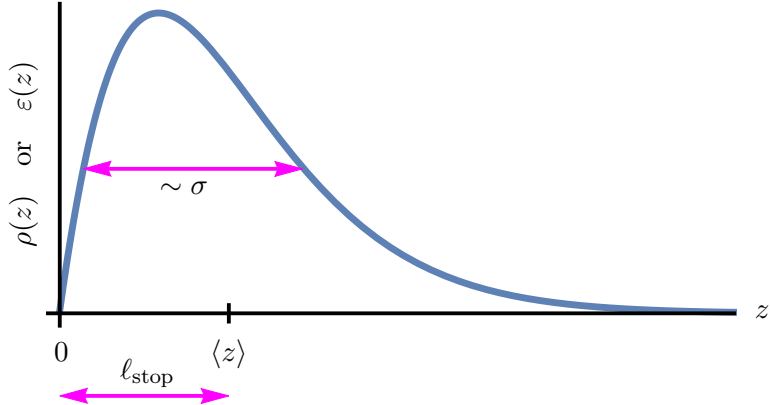}}
  \put(20,17){0}
  \put(79,17){$\langle z \rangle$}
  \put(285,33){$z$}
  \put(45,10){$\ell_{\rm stop}$}
  \put(60,80){$\sim \sigma$}
  \put(3,60){\rotatebox{90}{$\rho(z)$ ~~or~~ $\varepsilon(z)$}}
  \end{picture}
  \caption{
     \label{fig:deposit}
     The stopping distance and width characteristic of where charge
     or energy is deposited by a shower.
     (The above curve is presented here for qualitative purposes,
     but the shape depicted just happens to be the precise
     {\it leading}-order numerical
     result for QED charge deposition $\rho(z)$ ---
     see appendix \ref{app:rhoNumeric}.  In that case, the specific
     ``$\sim\sigma$'' line shown above has length $2\sigma$.)
  }
\end {center}
\end {figure}

The average stopping distance $\lstop$
was the first moment (\ref{eq:lstopdef})
of the spatial
distribution $\rho(z)$ or $\varepsilon(z)$
of where charge or energy is deposited by the shower. 
But one could also ask after the width $\sigma$ of the region across
which those quantities are deposited, as depicted in
fig.\ \ref{fig:deposit}.  As we'll see later, both $\sigma$ and
$\lstop$ are of the same order:
\begin {equation}
  \sigma \sim \lstop \sim \frac{1}{\alpha}\sqrt{\frac{\hat q}{E}} .
\label {eq:sigmascale}
\end {equation}
So the scale of $\hat q$ will cancel if we consider their ratio
$\sigma/\lstop$.
We propose computing the dimensionless ratio,
\begin {equation}
   \frac{\sigma}{\ell_{\rm stop}}
   \equiv
   \frac{
           \bigl[ \langle z^2 \rangle - \langle z \rangle^2 \bigr]^{1/2}
        }{ \langle z \rangle }
   \,,
\label {eq:sigmadef}
\end {equation}
which could be taken for either
the deposited charge or deposited energy distributions,
similar to (\ref{eq:lstopdef}).
We will discuss how to compute the
formal expansion
of (\ref{eq:sigmadef}) in powers of $\alpha(Q_\perp)$, writing
\begin {equation}
   \frac{\sigma}{\ell_{\rm stop}} =
   \left(\frac{\sigma}{\ell_{\rm stop}}\right)^{\!\!(0)}
   \bigl[ 1 + \test + O(\alpha^2) \bigr] .
\label {eq:sigratio}
\end {equation}
Here, as in (\ref{eq:expansion1}), the superscript ``(0)'' indicates
an approximation based only on the results for single splitting rates,
ignoring overlapping formation length effects.
$\test$ represents the relative size of the effect of overlapping formation
lengths on $\sigma/\ell_{\rm stop}$,
calculated to first order in $\alpha$.  Our proposed test is whether or not
$\test$ is at least somewhat small compared to 1.
One may also look at dimensionless ratios
involving higher moments of the distributions, but for the moment we
focus on (\ref{eq:sigmadef}).

Note from (\ref{eq:sigmascale}) that the leading-order dependence on
$\alpha$ will also cancel in the ratio $\sigma/\lstop$.  Since the
leading-order term $(\sigma/\lstop)^{(0)}$ in (\ref{eq:sigratio}) is independent
of $\alpha=\alpha(\mu)$, there will be no explicit $\ln\mu$ dependence
in the NLO correction.  So there will be no explicit renormalization
scale ambiguity when we discuss the relative size $\test$ of the
NLO correction as a function of $\alpha$.  In contrast, we had to
deal with the ambiguity when discussing the relative size of the NLO
correction in the expansion (\ref{eq:expansion1}) of $\lstop$ because
there the leading-order result $\lstop^{(0)}$ depended on $\alpha$.

\begin {figure}[t]
\begin {center}
  \includegraphics[scale=1.0]{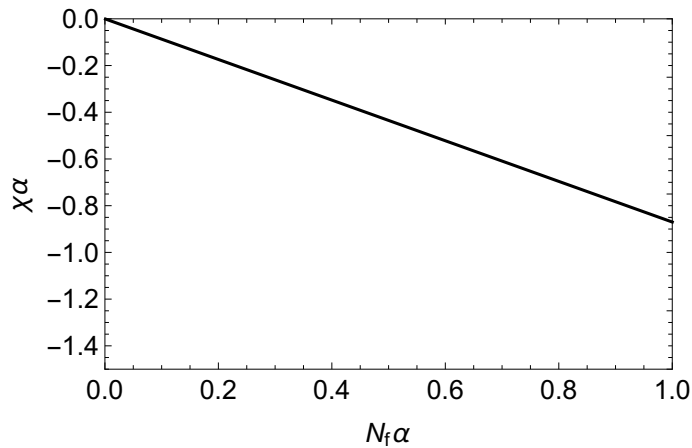}
  \caption{
     \label{fig:sigratioPlot}
     The relative size $\test$ of the NLO correction
     in large-$\Nf$ QED to
     the ratio $\sigma/\lstop$ for the deposited charge distribution.
     The horizontal axis is $\Nf\alpha$, where it is not necessary
     for this ratio
     at this order to distinguish the precise choice of renormalization
     scale (or scheme) other than that $\mu \sim (\hat q E)^{1/4}$
     parametrically.
  }
\end {center}
\end {figure}

As a concrete example,
a preview of our results in large-$\Nf$ QED of the relative size
$\test$ of the NLO correction to $\sigma/\lstop$ is
\begin {equation}
   \test = -0.870 \, \Nf\alphaqed .
\end {equation}
We plot this in fig.\ \ref{fig:sigratioPlot} for the sake of visual
comparison to fig.\ \ref{fig:lstopPlot}.  The two measures give roughly
similar conclusions for large-$\Nf$ QED: For $\Nf\alpha \simeq 1$,
overlap corrections are large, in which case the weakly-coupled splitting
picture (as in fig.\ \ref{fig:cartoon})
of the high-energy particles in showers would not be a very good picture.
In contrast, for $\Nf\alpha \le 0.2$,
the overlap corrections are only a modest effect.

It was of course a foregone conclusion that
{\it parametrically} the weakly-coupled splitting picture would break
down at $\Nf\alpha \sim 1$, but it was not previously clear whether
quantitatively that meant $\Nf\alpha \simeq 1$ or $\Nf\alpha/2\pi \simeq 1$ or
something else.


\subsection{Why not talk about \boldmath$dE/dz$?}

It is more common and traditional to package discussions of energy loss into
the evaluation of the rate $dE/dz$ of energy loss of a particular particle per
unit length of medium.  So why have we instead been focusing on stopping
lengths and the charge deposition distribution?
As we'll explain in section \ref{sec:MoreAssumptions},
the definition of $dE/dz$ becomes
ambiguous, in the general case, once one considers
corrections from overlapping formation times.
In this paper, our philosophy is to consider tests of the importance of
overlapping formation times which can be applied in general situations.

\begin {figure}[t]
\begin {center}
  \includegraphics[scale=1.0]{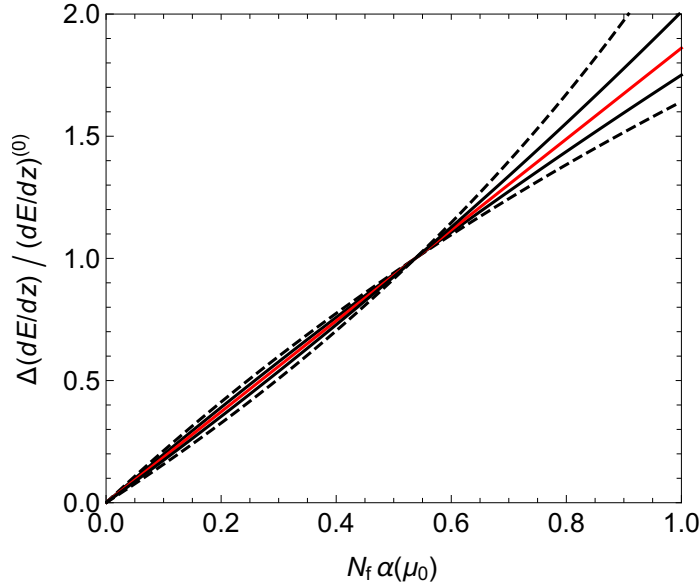}
  \caption{
     \label{fig:dEdzPlot}
     Like fig.\ \ref{fig:lstopPlot} but here for the relative size
     $\Delta(dE/dz)/(dE/dz)^{(0)}$ of NLO corrections to the
     electron energy loss rate.
  }
\end {center}
\end {figure}

That said, it will turn out that $dE/dz$ {\it is} unambiguously
defined for the example we use for numerics in this paper:
electron energy loss in large-$\Nf$ QED.  (It is also well defined
for quark energy loss in large-$\Nc$ QCD.)  So for this case, we can
make contact with the more traditional $dE/dz$ by showing in
figure \ref{fig:dEdzPlot} a plot of the relative size
$\Delta(dE/dz)/(dE/dz)^{(0)}$ of the NLO correction to $dE/dz$,
analogous to our plot in fig.\ \ref{fig:lstopPlot} for
$\Delta\lstop/\lstop^{(0)}$.

\subsection{Outline}

In the next section, we first state a little more clearly some of our
assumptions.  Then we present in the main text the simplest derivations
(that will be adequate for the explicit case of large-$\Nf$ QED analyzed
in this paper)
of the moments $\langle z^n \rangle$ of the charge deposition distribution
$\rho(z)$.  From those results, we derive explicit integral formulas for
$\Delta\lstop/\lstop$ and the measure $\test$ in terms of
leading-order and NLO contributions to differential rates $d\Gamma/dx$
for splitting.
We leave discussion of energy stopping distances, and more
general techniques that can handle QCD double logarithms, to appendices.
In section \ref{sec:QED}, we then discuss in more detail the origin
of the leading-order and NLO contributions to the relevant
rates $d\Gamma/dx$, making contact with the explicit results and
expressions given for large-$\Nf$ QED rates derived in ref.\ \cite{QEDnf},
which are then used to produce our final numerical results of
figs.\ \ref{fig:lstopPlot} and \ref{fig:sigratioPlot}.
The NLO contributions to the splitting rates contain both
(i) overlap corrections
$\Delta\Gamma/dx\,dy$ for two consecutive emissions (like the overlap
shown in fig.\ \ref{fig:shower}b), and (ii) corresponding NLO virtual
corrections to leading-order, single-splitting results.
At the moment, (ii) has not yet been computed for
(large-$\Nc$) QCD but has been for large-$\Nf$ QED.
In section \ref{sec:RenormScale}, we discuss
more about renormalization scale dependence and discuss how the breakdown
of QED at high momentum scales (the Landau pole) is not a concern even
for the relatively large values of $\Nf\alphaqed$ we have investigated.

In large part, we motivated consideration of
the relative size $\test$ of the NLO correction to
$\sigma/\lstop$ (fig.\ \ref{fig:sigratioPlot}) by looking for
tests of strong vs.\ weak splitting
that would be insensitive to corrections to $\hat q$.
Section \ref{sec:caveats} discusses why
we anticipate that, for QCD, this particular test will be only
partly successful at eliminating sensitivity to $\hat q_{\rm eff}$.
Finally, we offer our conclusions in section \ref{sec:conclusion}.


\section{Stopping distributions}
\label {sec:stop0}

\subsection {Simplifying assumptions}

For simplicity, both for the sake of calculation and the sake of
constructing as simple a thought experiment as possible to test
weakly-coupled vs.\ strongly-coupled splitting, we will treat
the underlying value of $\hat q$ as constant in space and time.
In contrast, the properties of
quark-gluon plasmas produced in actual heavy-ion collisions depend
on both.

In quoting results for large-$\Nf$ QED, we will make a further
simplification.  This theory does not have the NLO double-log enhancements
(\ref{eq:dbllog0L}), but that does not mean that the value of
$\hat q$ relevant to splitting calculations is independent of formation
length and so of energy.  For QED, there is already (single) logarithmic
dependence in the value of leading-order $\hat q^{(0)}$.
(For a very brief review
qualitatively comparing and contrasting
the QED and QCD energy dependence of $\hat q$ in
the language of this paper, see, for example, Appendix C of
ref.\ \cite{QEDnf}.)
Since our interest is not so much in the details of large-$\Nf$ QED but
in using it as an example for the calculation of overlap corrections,
we will bypass this particular complication
and simply treat $\hat q^{(0)}$ as independent of energy.
One could do better if interested: For QED (unlike QCD)
the tentative assumption
$\Nf\alphaqed\bigl( (\hat q E)^{1/4} ) \ll 1$ of weak splitting
in our calculation at large energy means that the QED plasma itself
is {\it also} weakly coupled, $\Nf\alphaqed(g T) \ll 1$.
So, if desired, one could use a
{\it perturbative} calculation of
the relevant $\hat q^{(0)}$ and its scale dependence.
Since this is unnecessarily specific to the QED case, we've chosen to
avoid complicating our presentation and will treat
$\hat q^{(0)}$ as constant.


\subsection {More assumptions for the analysis in the main text}
\label {sec:MoreAssumptions}

In the main text, we will focus on charge deposition rather than energy
deposition because it is simpler to discuss at NLO.  We leave discussion
of energy deposition to appendix \ref{app:energy}.

\begin {figure}[t]
\begin {center}
  \includegraphics[scale=0.5]{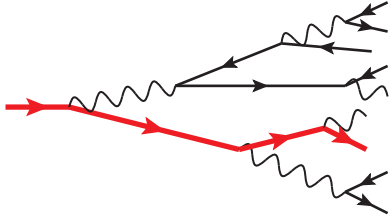}
  \caption{
     \label{fig:follow}
     The red (heavy) line above shows an example of following an
     initial fermion through a shower to see where it is deposited.
     When the NLO effect of overlapping formation times is
     considered, this picture is
     only valid in certain theories like large-$\Nf$ QED (see text).
  }
\end {center}
\end {figure}

\begin {figure}[t]
\begin {center}
  \includegraphics[scale=0.35]{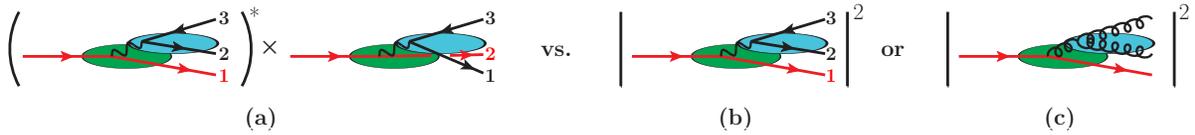}
  \caption{
     \label{fig:ambiguity}
     For overlapping formation time contributions to the splitting of
     a fermion, a comparison of contributions to the rate where
     identification of the direct heir of the parent fermion
     (a) is or (b,c) is not ambiguous.  (a,b) apply to both QED and
     QCD (with the wavy line a photon or gluon respectively), whereas
     (c) applies only to QCD.  (b) dominates over (a) in large-$\Nf$
     QED, and (c) dominates over both (a) and (b) in large-$\Nc$ QCD.
  }
\end {center}
\end {figure}

We will also focus in the main text on the simplest derivations that
will get us results for our example of large-$\Nf$ QED, and leave
discussion of more complicated situations to the appendices.  In
particular, we focus here on situations where the net ``charge''
carried by the parent of each splitting is unambiguously carried by
just one, identifiable daughter of that splitting, with more general
discussions in appendix \ref{app:energy}.
Graphically, this corresponds to simply following the original fermion
line through the entire shower, such as the red line depicted in
fig.\ \ref{fig:follow}, and then asking for the probability
distribution for where that red line comes to a stop in the medium.
If the splittings do not overlap (e.g.\ as in a calculation that used
only leading-order results for splitting processes), there would be no
problem: we could unambiguously identify the red line of
fig.\ \ref{fig:follow}.  At next-to-leading order, however,
overlapping formation times allow the possibility of non-negligible
interference between different final-state fermions, as shown by
fig.\ \ref{fig:ambiguity}a.
However, this interference is suppressed in large $N$.
For both large-$\Nf$ QED and large-$\Nc$ QCD,
it is suppressed (after summing over final
fermion flavors or colors) compared to
fig.\ \ref{fig:ambiguity}b, for which the direct heir of
the original fermion is unambiguous: the chance
that the original fermion and the
photon/gluon-produced pair have the same flavor or color is $1/\Nf$ or
$1/\Nc$ suppressed,
and distinguishable fermions cannot be confused with one another.
In large-$\Nc$ QCD, both figs.\ \ref{fig:ambiguity}(a,b) are
suppressed compared to fig.\ \ref{fig:ambiguity}c after summing over
final colors, and fig.\ \ref{fig:ambiguity}c has
no ambiguity regarding which daughter carries the quark number
of the initial quark.

\begin {figure}[t]
\begin {center}
  \includegraphics[scale=0.5]{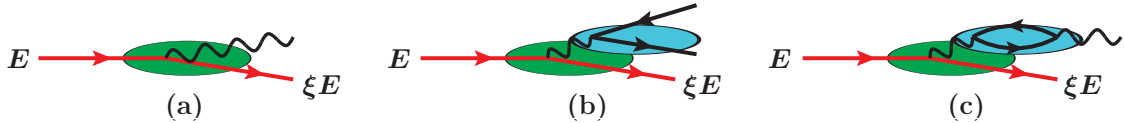}
  \caption{
     \label{fig:dGamma}
     Examples of processes which would contribute to the calculation
     of the differential rate $d\Gamma/d\xi$ at which a particle
     carrying a conserved charge (such as fermion number) degrades in
     energy:
     (a) leading-order bremsstrahlung;
     (b) the NLO effect of a subsequent splitting that has an
     overlapping formation time with the bremsstrahlung;
     (c) a virtual correction to the leading-order bremsstrahlung
     (with overlapping formation time).
     In the case of large-$\Nc$ QCD, the initial particle would be a quark
     (for study of charge deposition), but
     the particles in the real or virtual pair produced in (b) or (c)
     would be gluons.
  }
\end {center}
\end {figure}

The advantage of cases where we can make this assumption is that,
freed from having to follow anything but the single red line
through any given shower like fig.\ \ref{fig:follow},
all that is
needed to determine charge stopping can be packaged into
a single differential rate $d\Gamma/d\xi$ for that red-line particle
to reduce its longitudinal momentum by a factor of $\xi$.
In a purely leading-order analysis, this rate would represent
the bremsstrahlung process of fig.\ \ref{fig:dGamma}a.
For a NLO analysis, it would also include the
corrections to that rate due to overlapping formation time effects
with the next splitting, such as depicted in figs.\ \ref{fig:dGamma}b
and c.%
\footnote{
  For a very generic description of treating overlap corrections
  as contributions to rates, see the discussion in section I.A of
  ref.\ \cite{seq}.
}
The rate $d\Gamma/d\xi$ has the added convenience that many of
the contributions to overlapping double splitting cancel each
other in $d\Gamma/d\xi$, which simplifies computational work and
is one reason we focus on charge deposition instead of energy
deposition in this paper.  We'll discuss the cancellations more
concretely in section \ref{sec:QED}.
Here, we will just assume that the formula for $d\Gamma/d\xi$ is known
through next-to-leading order.

The ambiguity of following the red line in fig.\ \ref{fig:ambiguity}a
is also the reason that the fermion energy loss rate $dE/dz$ is ambiguous
(beyond leading order) outside of the large-$N$ limit.

In what follows, we first discuss a general equation (given the above
assumptions) whose solution determines the statistical distribution
of charge $\rho(z)$ in terms of splitting rates $d\Gamma/d\xi$.
After, we will discuss the fairly simple solution for the moments of
$\rho$ if one ignores logarithmic factors in rates.  We then present a trick
for extending that analysis to the case where next-to-leading-order
splitting rates
contain terms involving single logarithms of energy, which
will cover the case of large-$\Nf$ QED.  Finally, we mention (and then defer
to the appendices) a different and computationally more difficult
method that could be
used in the case of more-complicated dependence on energy, such as
double logarithms.


\subsection {Integro-differential equation for the charge
             distribution \boldmath$\rho(z)$}

Let $\rho(E,z)$ be the distribution of charge deposition from an initial
charge of energy $E$.
To find an equation for $\rho$ in terms of splitting rates, consider
$\rho(E,z{+}\Delta z)$ where $\Delta z$ is tiny, and think of traveling
the distance $z{+}\Delta z$ as first traveling $\Delta z$ followed
by traveling distance $z$.
In the first $\Delta z$, the particle has a chance of not
splitting at all, in which case the chance of traveling the remaining
distance $z$ will just be $\rho(E,z)$.  Alternatively, there is a chance
that the particle splits in the first $\Delta z$, reducing its energy
to $\xi E$, in which case the chance of traveling the remaining distance
is $\rho(\xi E,z)$.  In formulas,
\begin {equation}
  \rho(E,z{+}\Delta z)
  \simeq
  [1 - \Gamma(E)\,\Delta z] \, \rho(E,z)
  +
  \int_0^1 d\xi \> \frac{d\Gamma}{d\xi}(E,\xi) \, \Delta z \, \rho(\xi E,z)
  ,
\end {equation}
where $\Gamma(E)$ is the splitting rate for a particle with
energy $E$.  Re-arranging terms and taking the limit $\Delta z \to 0$,
\begin {equation}
  \frac{\partial \rho(E,z)}{\partial z}
  =
  - \Gamma(E) \, \rho(E,z)
  +
  \int_0^1 d\xi \> \frac{d\Gamma}{d\xi}(E,\xi) \, \rho(\xi E,z) ,
\label {eq:rhoeq0}
\end {equation}
which can be rewritten as
\begin {equation}
  \frac{\partial \rho(E,z)}{\partial z}
  =
  -
  \int_0^1 d\xi \> \frac{d\Gamma}{d\xi}(E,\xi)
  \bigl[ \rho(E,z) - \rho(\xi E,z) \bigr] .
\label {eq:rhoeq}
\end {equation}

The total LPM splitting rate $\Gamma = \int (d\Gamma/d\xi) d\xi$ is typically
infrared (IR)
divergent from soft bremsstrahlung (at leading order in QCD and from NLO overlap
corrections in large-$\Nf$ QED), corresponding to $\xi \to 1$ above.
However, the factor $\rho(E,z) - \rho(\xi E,z)$ in the integrand
of (\ref{eq:rhoeq}) will give an extra suppression as $\xi \to 1$ that
makes that $\xi$ integral finite.  So, even though we introduced IR-divergent
quantities in the derivation of (\ref{eq:rhoeq}), the final equation
is IR-safe.

The above argument for the $\rho$ equation, as well as what we will do
next, was inspired by related arguments made directly for
the first moment $\lstop$ of $\rho$ in ref.\ \cite{stop}.
It is also reminiscent of evolution equations explored in refs.\
\cite{Nevolve1,Nevolve2} for the
distribution $N(E,x,t)$ of shower particles in longitudinal
momentum fraction $x$ as a function of time $t{=}z$,
except that
here we focus directly on the spatial distribution $\rho$
of deposition and need not keep track of the detailed distribution
$N$ in $x$.%
\footnote{
  Refs.\ \cite{Nevolve1,Nevolve2,Nevolve3}
  studied $N(E,x,t)$ using leading-order
  splitting rates and found that they could make a great deal of
  interesting and useful progress in finding closed-form
  {\it analytic} results
  if they modified the leading-order LPM rate $d\Gamma/d\xi$ to
  a not-too-different mathematical function that agrees arbitrarily
  well as $\xi$ approaches 0 or 1 and differs a bit in the middle.
  We will not
  make that replacement in our work,
  and we will be including NLO corrections,
  which have a more complicated structure.
}


\subsection {Warm-up: charge stopping with no logarithms}
\label {sec:lstopNoLog}

\subsubsection {Results for moments}

Our simple parametric estimate
(\ref{eq:lrad1}) showed that splitting rates scale with energy
as $E^{-1/2}$.  If we ignore logarithmic corrections (e.g.\ from
energy dependence of $\alpha$ and $\hat q$), the simple energy
scaling allows us to derive simple results
for stopping lengths $\ell_{\rm stop}$,
and for other moments of the deposited charge distribution, in
terms of $d\Gamma/d\xi$.  To see this, pull out the explicit energy
dependence of the splitting rate by writing
\begin {equation}
  \frac{d\Gamma}{d\xi}(E,\xi)
  =
  E^{-\nu} \, \frac{d\tilde\Gamma}{d\xi}(\xi)
\label {eq:dGscale}
\end {equation}
where
\begin {equation}
  \nu = \tfrac12 .
\end {equation}
(The only reason we introduce the symbol $\nu$ for the power
$\tfrac12$ is that
later it will help in generalizing this analysis to the case of single
logarithms.)
Because the rate scales as $E^{-\nu}$, the distance that the charge travels
before stopping in the medium will scale as $E^{+\nu}$.
Let's factor this scaling out of the probability distribution
$\rho(E,z)$ by replacing it with a probability distribution
$\tilde\rho(\tilde z)$ in terms of $\tilde z \equiv z/E^\nu$,
normalized so that $\rho\,dz = \tilde\rho\,d\tilde z$.  That is,
\begin {equation}
  \rho(E,z) = E^{-\nu} \, \tilde\rho(E^{-\nu}z) .
\label {eq:rhoscale}
\end {equation}
Plugging (\ref{eq:dGscale}) and (\ref{eq:rhoscale}) into the
$\rho$ equation (\ref{eq:rhoeq}) gives
\begin {equation}
  \frac{\partial \tilde\rho(E^{-\nu} z)}{\partial z}
  =
  -
  \int_0^1 d\xi \> \frac{d\tilde\Gamma}{d\xi}
  \Bigl[ E^{-\nu} \, \tilde\rho(E^{-\nu} z)
         - (\xi E)^{-\nu} \, \tilde\rho\bigl((\xi E)^{-\nu}z\bigr) \Bigr] ,
\end {equation}
which can be rewritten as
\begin {equation}
  \frac{d \tilde\rho(\tilde z)}{d\tilde z}
  =
  -
  \int_0^1 d\xi \> \frac{d\tilde\Gamma}{d\xi}
  \bigl[ \tilde\rho(\tilde z)
         - \xi^{-\nu} \tilde\rho\bigl(\xi^{-\nu}\tilde z\bigr) \bigr] .
\label {eq:rhoeqscaled}
\end {equation}
As an aside, we note that numerical solution of this equation
for {\it leading}-order QED splitting was used to generate the
precise shape of
$\rho(z)$ shown as an example in fig.\ \ref{fig:deposit}
(see appendix \ref{app:rhoNumeric}).

Now multiply both sides by $\tilde z^n$ and integrate over $\tilde z$
to get (after integration by parts on the left-hand side of the
equation) a recursive relationship between the moments of the
probability distribution $\tilde\rho$:
\begin {equation}
  -n \langle\tilde z^{n-1}\rangle
  =
  -
  \int_0^1 d\xi \> \frac{d\tilde\Gamma}{d\xi}
  \bigl[ \langle \tilde z^n \rangle
         - (\xi^\nu)^n \langle \tilde z^n \rangle \bigr] ,
\end {equation}
giving
\begin {subequations}
\label {eq:recursionA}
\begin {equation}
   \langle \tilde z^n \rangle =
   \frac{ n \langle \tilde z^{\,n-1} \rangle }
        { \int_0^1 d\xi \> \frac{d\tilde\Gamma}{d\xi} [1-\xi^{n\nu}] }
   \,.
\label {eq:recursionA1}
\end {equation}
From now on, we will assume that the charge $q_0$ of the particle
initiating the shower is normalized to be $q_0{=}1$, so that
$\rho$ is normalized such that $\langle 1 \rangle = 1$.
The case $n{=}1$ of (\ref{eq:recursionA1}) then
gives the formula
\begin {equation}
  \lstop = \langle z \rangle = E^\nu \langle \tilde z \rangle
  = 
   \frac{ E^\nu }
        { \int_0^1 d\xi \> \frac{d\tilde\Gamma}{d\xi} [1-\xi^\nu] }
   \,,
\label {eq:lstopA}
\end {equation}
\end {subequations}
in agreement with ref.\ \cite{stop}.%
\footnote{
  Specifically (3.7--8) of ref.\ \cite{stop}, where the $1{-}x$ there
  is the $\xi$ here.
}
But the recursion relation (\ref{eq:recursionA1}) derived here
lets us find higher
moments like $\langle \tilde z^2 \rangle$ and so
$\sigma
  = E^\nu \sqrt{\langle \tilde z^2 \rangle - \langle \tilde z \rangle^2}$.


\subsubsection {Measures of size of NLO correction}

We'll now expand (\ref{eq:recursionA}) to first order in $\alpha$
to get formulas for the measures $\Delta\lstop/\lstop^{(0)}$ and $\test$
of the relative size of corrections to showering from overlapping formation
times.  To this end, decompose the splitting rate $d\Gamma/d\xi$ into
its leading order piece (the usual LPM calculation for single splitting)
and its NLO correction:
\begin {equation}
  \frac{d\Gamma}{d\xi}
  \simeq
  \frac{d\Gamma^{(0)}}{d\xi}
  +
  \frac{d\Gamma^{({\rm NLO})}}{d\xi} \,.
\label {eq:dGexpand}
\end {equation}
The NLO piece is suppressed by $O(\alpha)$ compared to the leading-order
rate, and the $\simeq$ above just indicates that we are ignoring
yet-higher order terms, suppressed by $O(\alpha^2)$ or more.

Plugging (\ref{eq:dGexpand}) into (\ref{eq:lstopA}) and expanding
$\lstop \simeq \lstop^{(0)} + \Delta\lstop$ gives
\begin {equation}
   \frac{\Delta\lstop}{\lstop^{(0)}}
   =
   -
   \frac{
     \int_0^1 d\xi \> \frac{d\tilde\Gamma^{({\rm NLO})}}{d\xi} [1-\sqrt{\xi}]
   }{
     \int_0^1 d\xi \> \frac{d\tilde\Gamma^{(0)}}{d\xi} [1-\sqrt{\xi}]
   } \,.
\end {equation}
We find the structure of later equations to be a little clearer if
we use the alternative notation
\begin {equation}
   \frac{\Delta\lstop}{\lstop^{(0)}}
   =
   -
   \frac{ \Avg[1-\sqrt{\xi}]^{({\rm NLO})} }
        { \Avg[1-\sqrt{\xi}]^{(0)} }
   \,,
\label {eq:lstopRatioA}
\end {equation}
where we define leading and next-to-leading order rate-weighted averages
of any function $f(\xi)$ as
\begin {equation}
   \Avg[f(\xi)]^{(0)} \equiv
     \int_0^1 d\xi \> \frac{d\tilde\Gamma^{(0)}}{d\xi} \, f(\xi) ,
   \qquad
   \Avg[f(\xi)]^{({\rm NLO})} \equiv
     \int_0^1 d\xi \> \frac{d\tilde\Gamma^{({\rm NLO})}}{d\xi} \, f(\xi) .
\end {equation}

Similarly, plugging (\ref{eq:dGexpand}) into the $n{=}2$ case
\begin {equation}
   \langle \tilde z^2 \rangle =
   \frac{ 2 \tildelstop }
        { \int_0^1 d\xi \> \frac{d\tilde\Gamma}{d\xi} [1-\xi] }
\end {equation}
of (\ref{eq:recursionA1}) gives
\begin {equation}
   \frac{\sigma}{\lstop}
   = \frac{\tilde\sigma}{\tildelstop}
   = \sqrt{ \frac{\langle\tilde z^2\rangle - \tildelstop^2}{\tildelstop^2} }
   \simeq
   \left(\frac{\sigma}{\lstop}\right)^{\!\!(0)}
   [ 1 + \test ]
\end {equation}
with
\begin {equation}
  \test =
   \frac{ \Avg[(1-\sqrt{\xi})^2]^{({\rm NLO})} }
        { 2 \Avg[(1-\sqrt{\xi})^2]^{(0)} }
   -
   \frac{ \Avg[1-\xi]^{({\rm NLO})} }
        { 2 \Avg[1-\xi]^{(0)} }
  \,.
\label {eq:testA}
\end {equation}

A feature of (\ref{eq:testA}) worth noting is that if
the next-to-leading order rate $d\Gamma^{(\rm NLO)}/d\xi$ contains
a piece that is any $\xi$-independent constant times the leading-order
rate $d\Gamma^{(0)}/d\xi$, that piece will completely cancel between
the two terms of (\ref{eq:testA}) and so will not contribute at all
to $\test$.  So, for instance, if the NLO rate were to contain contributions
corresponding to $\xi$-independent NLO corrections to the $\hat q$
appearing in the leading-order rate, such corrections to $\hat q$ would
not contribute to $\test$.  This was part of our earlier motivation to
consider the ratio $\sigma/\lstop$ and thence $\test$.


\subsection {Charge stopping with single logarithms}
\label {sec:lstop1Log}

We now generalize the previous analysis to a case that will
cover QED in the approximations used in this paper.
Consider the parametric formula (\ref{eq:lrad1}) for the distance
$l_{\rm rad}$ between nearly-democratic splittings.  Taking its
inverse, the correspond splitting rate is
\begin {equation}
   \Gamma \sim \alpha \sqrt{\frac{\hat q}{E}} \qquad
   (\mbox{nearly-democratic splittings}).
\label {eq:Gdem}
\end {equation}
In general, the {\it leading-order} splitting rate $(d\Gamma/d\xi)^{(0)}$
is proportional to the $\alpha = \alpha(\mu)$
associated with the splitting vertex, like in the parametric formula above.
As previously discussed, the renormalization and running of $\alpha$ means
that logarithms $\ln(\mu/Q_\perp)$ must therefore appear at next-to-leading
order, similar to (\ref{eq:lstopRatio0qed}).  Specifically, for
large-$\Nf$ QED, ref.\ \cite{QEDnf} found by explicit calculation that
the rate (\ref{eq:dGexpand}) has leading-order behavior
$d\Gamma^{(0)}/d\xi \propto E^{-1/2}$, like (\ref{eq:dGscale}) and
(\ref{eq:Gdem}), but NLO corrections of the form%
\footnote{
  Our choice for how to write the argument of the explicit logarithm in
  (\ref{eq:NLOlog}) does not mean that the appropriate scale for
  $\alpha(Q_\perp)$ is $(\hat q E)^{1/4}$ [the nearly-democratic splitting
  estimate] for any value of $\xi$.  But since
  $\ln(\mu/[f(\xi)\,(\hat q E)^{1/4}])$ can be written as
  $\ln(\mu/(\hat q E)^{1/4}) - \ln\bigl(f(\xi)\bigr)$ for any $\xi$-dependence
  $f(\xi)$, the $\xi$ dependence of the scale can be absorbed into the
  last term of (\ref{eq:NLOlog}).  Here, the adjustments we need to make to
  our previous derivation of formulas for moments of the stopping distribution
  will only revolve around the dependence of the argument of the logarithm
  on $E$, not on $\xi$.
}
\begin {align}
   \frac{d\Gamma^{({\rm NLO})}}{d\xi}
   &=
   - \beta_0 \alphaqed \, \frac{d\Gamma^{(0)}}{d\xi} \,
       \ln\Bigl( \frac{\mu}{(\hat q E)^{1/4}} \Bigr)
\nonumber\\ & \qquad
   + E^{-1/2} \times (\mbox{something independent of $E$}) ,
\label {eq:NLOlog}
\end {align}
where here
\begin {equation}
   \beta_0 = \frac{2\Nf}{3\pi} \qquad \mbox{(QED)}
\end {equation}
is the coefficient in the 1-loop renormalization group $\beta$-function
for $\alphaqed$.
We'll find it convenient to rewrite (\ref{eq:NLOlog}) as
\begin {align}
   \frac{d\Gamma^{({\rm NLO})}}{d\xi}
   &=
   \tfrac14 \beta_0 \alphaqed \, \frac{d\Gamma^{(0)}}{d\xi} \,
       \ln E
\nonumber\\ & \qquad
   + E^{-1/2} \times (\mbox{something independent of $E$}) .
\label {eq:NLOlog1}
\end {align}

In this section, we
consider generally cases
like (\ref{eq:NLOlog1}) where the NLO corrections to the leading-order
rate may have, in addition to terms with energy dependence $E^{-1/2}$,
a term with energy dependence $E^{-1/2} \ln E$ proportional to the
leading-order rate.  
Similar to (\ref{eq:dGscale}), we will scale out all the explicit factors
of $E^{-1/2}$ by writing
\begin {equation}
  \frac{d\Gamma}{d\xi}(E,\xi)
  =
  E^{-1/2} \frac{d\tilde\Gamma}{d\xi}(E,\xi)
  \simeq
  E^{-1/2} \Biggl\{
    \frac{d\tilde\Gamma^{(0)}}{d\xi} (\xi)
    + \tfrac14 \beta_0 \alpha \, \frac{d\tilde\Gamma^{(0)}}{d\xi} (\xi) \, \ln E
    + \frac{d\tilde\Gamma^{({\rm NLO}')}}{d\xi} (\xi)
  \Biggr\} ,
\label {eq:NLOlog2}
\end {equation}
where ${\rm NLO}'$ means all the NLO terms except the $\ln E$ term.
In principle, the coefficient ``$\tfrac14 \beta_0$'' above
could represent any coefficient of any sort of
NLO single-log correction to the leading-order result, but we
will use the notation $\tfrac14\beta_0$ relevant to the QED application
(\ref{eq:NLOlog1}).

We can now use the following trick.  Through next-to-leading order,
(\ref{eq:NLOlog2}) is equivalent to
\begin {equation}
  \frac{d\Gamma}{d\xi}(E,\xi)
  \simeq
  E^{-\nu} \Biggl\{
    \frac{d\tilde\Gamma^{(0)}}{d\xi} (\xi)
    + \frac{d\tilde\Gamma^{({\rm NLO}')}}{d\xi} (\xi)
  \Biggr\}
  \equiv
  E^{-\nu} \,
    \frac{d\hat\Gamma}{d\xi} (\xi)
\label {eq:NLOlog3}
\end {equation}
with
\begin {equation}
   \nu = \tfrac12 - \tfrac14 \beta_0 \alpha .
\label {eq:nu}
\end {equation}
This allows us to use the same formulas (\ref{eq:recursionA}) that
we derived for moments in the previous section, except that $\nu$
is now (\ref{eq:nu}).  Accounting for this difference in $\nu$ when
we expand in $\alpha$, (\ref{eq:lstopRatioA}) and (\ref{eq:testA}) are
modified to
\begin {equation}
   \frac{\Delta\lstop}{\lstop^{(0)}}
   =
   -
   \frac{ \Avg[1-\sqrt{\xi}]^{({\rm NLO})} }
        { \Avg[1-\sqrt{\xi}]^{(0)} }
   -
   \frac{ \beta_0 \alpha \Avg[\sqrt{\xi}\ln\xi]^{(0)} }
        { 4 \Avg[1-\sqrt{\xi}]^{(0)} }
\label {eq:lstopRatioB}
\end {equation}
and
\begin {multline}
  \test =
   \frac{ \Avg[(1-\sqrt{\xi})^2]^{({\rm NLO'})} }
        { 2 \Avg[(1-\sqrt{\xi})^2]^{(0)} }
   -
   \frac{ \Avg[1-\xi]^{({\rm NLO'})} }
        { 2 \Avg[1-\xi]^{(0)} }
\\
   +
   \frac{ \beta_0 \alpha \Avg[(\sqrt\xi-\xi)\ln\xi]^{(0)} }
        { 4 \Avg[(1-\sqrt{\xi})^2]^{(0)} }
   -
   \frac{ \beta_0 \alpha \Avg[\xi\ln\xi]^{(0)} }
        { 4 \Avg[1-\xi]^{(0)} }
   \,.
\label {eq:testB}
\end {multline}
These are the formulas that we will use later with large-$\Nf$
QED results
for $d\Gamma/d\xi$ in order to produce figs.\ \ref{fig:lstopPlot}
and \ref{fig:sigratioPlot}.


\subsection {Charge stopping with double logarithms, etc.}
\label {sec:dbllog}

For cases (such as QCD) where the NLO contribution to $d\Gamma/d\xi$
contains terms with more complicated energy dependence than $E^{-1/2}
\ln E$, we have not figured out a way to get formulas as simple as
(\ref{eq:lstopRatioB}) and (\ref{eq:testB}) for our tests
of overlap corrections to shower development.  One could
turn to direct numerical simulation of the original equation
(\ref{eq:rhoeq}) for $\rho(E,z)$, but in the general case we have not
so far figured out any particularly efficient way (numerically or
otherwise) to solve the equation and cleanly isolate the expansion of
that solution to first order in $\alpha$.
We present a different approach in Appendix \ref{app:dbllog},
which has been relegated to an appendix because (i) it is more
complicated to derive and will be more complicated to implement
numerically, and (ii) the previous results (\ref{eq:lstopRatioB}) and
(\ref{eq:testB}) are all that we will need for the large-$\Nf$ QED
results presented in this paper.  Also, since we do not
yet have full QCD results for $d\Gamma/d\xi$, we leave the
task of numerically implementing the procedure described in
Appendix \ref{app:dbllog} to later work.

Note that, because all our measures of the size of NLO corrections
are linear in $d\Gamma^{\rm(NLO)}/d\xi$, one could use different techniques
for different contributions to $d\Gamma^{\rm(NLO)}/d\xi$ and then sum
together the results.  So, if desired,
(\ref{eq:lstopRatioB}) and (\ref{eq:testB}) could be used to
calculate the contribution to $\Delta\lstop/\lstop$ and $\test$ from
all terms of $d\Gamma^{\rm(NLO)}/d\xi$ {\it except} the $\ln^2 E$ term,
and then just the $\ln^2 E$ contribution could be evaluated using
some other method like Appendix \ref{app:dbllog}.


\section{The example of large-\boldmath$\Nf$ QED}
\label {sec:QED}

\subsection {The relevant diagrams}

\begin {figure}[tp]
\begin {center}
  \includegraphics[scale=0.43]{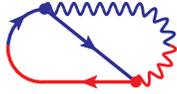}
  \caption{
     \label{fig:diagLO}
     Leading-order time-ordered diagram for $e \to \gamma e$.
     Blue represents a contribution to
     the amplitude and red represents a contribution to the conjugate
     amplitude.  Repeated
     interactions with the medium are present but not explicitly shown.
     This diagram should be added to its complex conjugate by taking
     $2\Re[\cdots]$.
  }
\end {center}
\end {figure}

\begin {figure}[tp]
\begin {center}
  \includegraphics[scale=0.43]{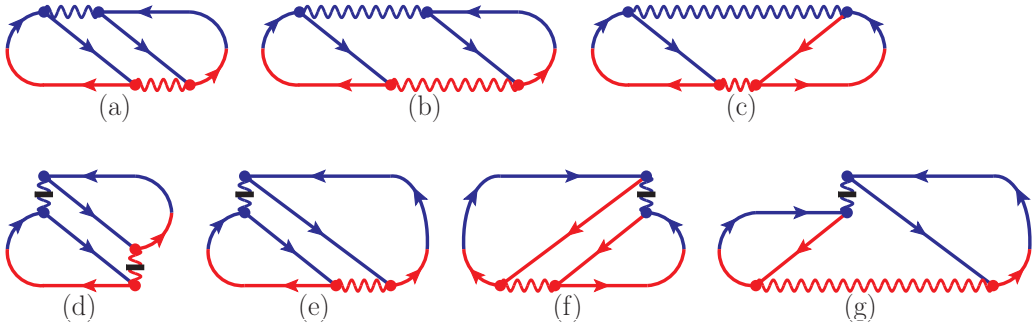}
  \caption{
     \label{fig:diags}
     Time-ordered interference diagrams for
     $e \to e \bar e e$ in large-$\Nf$ QED \cite{QEDnf}.
     Complex conjugates should also be
     included by taking $2\Re[\cdots]$ of the above.
  }
\end {center}
\end {figure}

\begin {figure}[tp]
\begin {center}
  \includegraphics[scale=0.43]{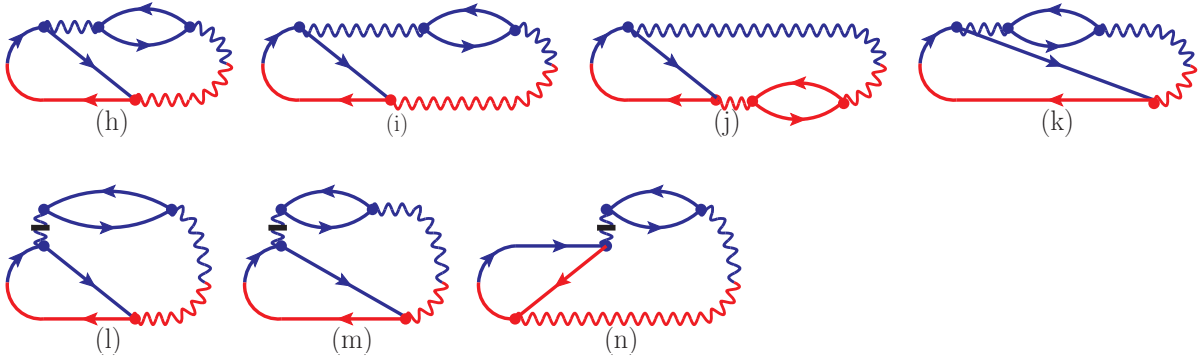}
  \caption{
     \label{fig:diagsVIRT}
     Time-ordered interference diagrams for the virtual correction to
     $e \to \gamma e$ in large-$\Nf$ QED \cite{QEDnf}.
     Again, complex conjugates should be included by taking $2\Re[\cdots]$.
  }
\end {center}
\end {figure}

In preceding sections, we have talked abstractly about the rate
$d\Gamma/d\xi$ relevant to following the original charged particle
through the medium in cases (like large $N$) where the original particle's
heir is always identifiable, even at next-to-leading order.
Here we will relate this $d\Gamma/d\xi$, for large-$\Nf$ QED,
to the specific diagrams and results of ref.\ \cite{QEDnf}.
In the conventions of ref.\ \cite{QEDnf}, the
leading-order rate $d\Gamma^{(0)}/d\xi$ is given by the
time-ordered interference diagram fig.\ \ref{fig:diagLO} (plus its
complex conjugate) for $e \to \gamma e$; the NLO corrections due
to overlapping real-double splitting $e \to \gamma e \to e \bar e e$ are
given by fig.\ \ref{fig:diags} (plus complex conjugates);
and the NLO virtual corrections to
single splitting are given by fig.\ \ref{fig:diagsVIRT}
(plus complex conjugates).
Each diagram represents a contribution to the rate for the process,
which is the product of an amplitude (represented by the blue part) times
a conjugate amplitude (represented by the red part).
In each diagram, only the high-energy particles are explicitly shown,
but each high-energy particle is interacting an arbitrary number of times
with the medium, and a medium average is performed.
See refs.\ \cite{2brem,seq,QEDnf} for more detail.
The vertical photon lines with bars across them represent photons with
longitudinal polarization in light-cone gauge, which produce
instantaneous interactions in light-cone time, whereas the unbarred photon
lines represent transverse photons.

All together, the $d\Gamma/d\xi$ discussed earlier in this paper is
\begin {subequations}
\label {eq:GammaSum}
\begin {equation}
  \frac{d\Gamma}{d\xi}
  \simeq
  \frac{d\Gamma^{(0)}}{d\xi}
  +
  \frac{d\Gamma^{\rm(NLO)}}{d\xi}
  \,,
\end {equation}
where the first term corresponds to fig.\ \ref{fig:diagLO} and where
\begin {align}
  \frac{d\Gamma^{\rm(NLO)}}{d\xi}
  &=
  \left(
  \int_0^1 d\yfrake
    \left[ \frac{\Delta \, d\Gamma}{d\xi \, d\yfrake} \right]_{e \to e\bar e e}
  \right)
  +
  \left[ \frac{\Delta \, d\Gamma}{d\xi} \right]^{\rm(NLO)}_{e\to\gamma e}
\nonumber\\
  &=
  2 \Re
  \left(
  \int_0^1 d\yfrake
    \left[ \frac{\Delta \, d\Gamma}{d\xi \, d\yfrake}
    \right]_{\mbox{\scriptsize(a--g)}}
  \right)
  +
  2 \Re
  \left[ \frac{\Delta \, d\Gamma}{d\xi} \right]_{\mbox{\scriptsize(h--n)}}
  \,.
\label {eq:GammaNLOsum}
\end {align}
\end {subequations}
In the last equation, the two terms correspond respectively to
figs.\ \ref{fig:diags} and \ref{fig:diagsVIRT}.
$\xi$ is called $\xe$ in ref.\ \cite{QEDnf} and is the factor by which the
original charge loses energy via single or overlapping double splitting.
$\yfrake$ represents the momentum fraction of the pair-produced electron
relative to the photon that produced it.  The $\Delta$ in $\Delta \, d\Gamma$
indicates that one should subtract from fig.\ \ref{fig:diags} for
$e \to e\bar e e$ the
result one would get instead if one simply combined the {\it leading}-order
formulas for the rate of $e \to \gamma e$ with that for $\gamma \to \bar e e$.
In this way, double counting is avoided when using
$d\Gamma/d\xi$ to generate shower development, and $\Delta\,d\Gamma$
represents the {\it corrections} to the rate due to overlapping formation
times.  See section I.A of ref.\ \cite{seq} for more explanation.
The $\Delta\,d\Gamma$ in the second term of (\ref{eq:GammaNLOsum}) indicates
a similar subtraction \cite{QEDnf} for fig.\ \ref{fig:diagsVIRT}, where the
second splitting is virtual.

Ref.\ \cite{QEDnf} shows that there is a simple relationship
(called there a back-end transformation) that relates the
double-splitting diagrams (a,b,c,e,g) to the virtual diagrams
(h,i,j,l,n) respectively.
For the particular combination (\ref{eq:GammaNLOsum}) [in which the
double-splitting diagrams are simply integrated over $\yfrake$ without
any other factors], these diagrams in fact cancel each other and so do
not need to be evaluated.
Another relationship was found between (m) and (e), in a way that we have
since discovered is related to (f) as%
\footnote{
  One finds this relationship if one simply writes out the expressions
  for diagrams (m) and (f) or evaluates the two diagrams numerically using
  the formulas of Ref.\ \cite{QEDnf}.
}
\begin {equation}
  2\Re
  \left[ \frac{d\Gamma}{d\xi} \right]_{\mbox{\scriptsize(m)}}
  =
  2\Re
  \int_0^1 d\yfrake
  \left[ \frac{d\Gamma}{d\xi\,d\yfrake} \right]_{\mbox{\scriptsize(f)}}
  .
\end {equation}
Putting all these cancellations and relations together,
(\ref{eq:GammaNLOsum}) can be simplified to
\begin {equation}
  \frac{d\Gamma^{\rm(NLO)}}{d\xi}
  =
  2 \Re
  \left[ \frac{d\Gamma}{d\xi} \right]_{\mbox{\scriptsize(k)}}
  +
  2 \Re
  \int_0^1 d\yfrake 
  \left(
    \left[ \frac{d\Gamma}{d\xi \, d\yfrake}
    \right]_{\mbox{\scriptsize(d)}}
    + 2 \left[ \frac{d\Gamma}{d\xi \, d\yfrake}
    \right]_{\mbox{\scriptsize(f)}}
  \right) .
\label {eq:GammaNLOsum2}
\end {equation}
The ``$\Delta$'' designation in $\Delta\,d\Gamma$ has been dropped
because the diagrams where that distinction was important
(diagrams a--c \cite{seq} and the related virtual diagrams) have canceled.
In consequence, for calculations of the charge stopping distance in
large-$\Nf$ QED, we need evaluate just three of the NLO
interference diagrams: the double-emission diagrams (d) and (f) and
the virtual correction diagram (k).


\subsection {Formulas and Numerical Integrals}

Formulas for relevant diagrams can be found in Appendix A of
ref.\ \cite{QEDnf}.  There we use variables $\xe$ and
$\ye$ to parametrize the momentum
fractions of daughters, where the relation to the $\xi$ and $\yfrake$
used above is
\begin {equation}
  \xe \equiv \xi, \qquad \ye \equiv (1-\xe) \yfrake .
\end {equation}
$\ye$ is the momentum fraction of the pair-produced electron relative to
the initial fermion of the double-splitting or virtual process, rather
than relative to the photon.
The leading-order rate is%
\footnote{
  The leading-order
  rate formula (\ref{eq:LOformula})
  is equivalent to the 1956 result by Migdal \cite{Migdal}
  if the
  latter is packaged into more modern, general notation.
  (See, for example, appendix C.4 of ref.\ \cite{QEDnf} for a translation.)
  The QCD analogs, not shown here, trace back to BDMPS \cite{BDMPS12,BDMPS3}
  and Zakharov \cite{Zakharov}.
}
\begin {subequations}
\label {eq:LOformula}
\begin {equation}
   \frac{d\Gamma^{(0)}}{d\xe} \equiv
   \left[\frac{d \Gamma}{d\xe} \right]_{\rm LO}
   = 2\Re \left[ \frac{d \Gamma}{d\xe} \right]_{x\bar x}
\end {equation}
with
\begin {equation}
  \left[ \frac{d \Gamma}{d\xe} \right]_{x\bar x}
  = \frac{\alphaqed}{2\pi} \, P_{e\to e}(\xe) \, i\Omega_0 ,
\end {equation}
\begin {equation}
  P_{e\to e}(\xe) = \frac{1 + \xe^2}{1-\xe} ,
  \qquad
  \Omega_0
   = \sqrt{ \frac{-i(1-\xe)\hat q}{2 \xe E} } \,.
\end {equation}
\end {subequations}

For next-to-leading order, formulas for the differential rates
in (\ref{eq:GammaNLOsum2}) above are given by
ref.\ \cite{QEDnf} eqs. (A33), (A35), and (A41).
The rate for the virtual diagram has the form
\begin {multline}
  \left[ \frac{d\Gamma}{d\xe} \right]_{\mbox{\scriptsize(k)}}
  \equiv
  \left[ \frac{d\Gamma}{d\xe} \right]_{x y y \bar x}^{\rm(ren)}
  =
  -
  \frac{\Nf\alphaqed}{3\pi}
  \left[ \frac{d\Gamma}{d\xe} \right]_{x\bar x} 
  \biggl(
    \ln\Bigl( \frac{\mu^2}{(1{-}\xe)E\Omega_\ix} \Bigr)
    + \gammaE
    - 2\ln2
    + \tfrac53
  \biggr)
\\
  +
  \int_0^{1-\xe} d\ye \>
  \left[ \frac{d\Gamma}{d\xe\,d\ye} \right]_{xyy\bar x}^{(\rm subtracted)} ,
\label {eq:dGk}
\end {multline}
involving an integral over the loop momentum fraction $\ye$,
with $[ d\Gamma/d\xe\,d\ye ]_{xyy\bar x}^{(\rm subtracted)}$
given by ref.\ \cite{QEDnf} eq.\ (A43).
Renormalization of diagram (k) has been carried out in the MS-bar scheme
($\MSbar$),
and so $\alphaqed = \alphaqed(\mu)$ is the MS-bar renormalized coupling.
Most of the differential rate formulas involves integrals
\begin {equation}
  \frac{d\Gamma}{d\xe\,d\ye} \propto
  \int_0^\infty d(\Delta t) \>
  D(\xe,\ye,\Delta t\bigr)
\label{eq:dGform}
\end {equation}
of complicated functions $D(\xe,\ye,\Delta t)$.
These integrals must be done numerically because we have failed in
most cases to
do them analytically.
Subsequently, an integral over $\ye$ or equivalently $\yfrake$
must be performed in (\ref{eq:GammaNLOsum}) and (\ref{eq:dGk}) to
obtain $d\Gamma/d\xi$.  Finally, integrals over $\xi$ must be
done to test NLO effects on shower development,
such as (\ref{eq:lstopRatioB}) and (\ref{eq:testB}).
All told, that's a triple numerical integral $(\xi,\yfrake,\Delta t)$
[equivalently $(\xe,\ye,\Delta t)$] of a complicated integrand.

We found even the initial integration (\ref{eq:dGform})
over $\Delta t$ to be numerically
expensive for $[ d\Gamma/d\xe\,d\ye ]_{xyy\bar x}^{(\rm subtracted)}$.
Presumably, more efficient numerical
methods could be developed, but we took the following approach.  We
evaluated $[ d\Gamma/d\xe\,d\ye ]_{xyy\bar x}^{(\rm subtracted)}$
for an appropriately chosen mesh of
points in the $(\xi,\yfrake)$ plane.  The function has integrable divergences
in various limits such as $\xi{\to}0$ or $1$ and/or $\yfrake{\to}0$.
Accounting for these divergences,
we then figured out a way to numerically
construct an interpolating function
throughout the $(\xi,\yfrake)$ integration region.
Using that interpolation, we then performed the final integration
over $(\xi,\yfrake)$ by brute force with standard integration software
(Mathematica \cite{Mathematica}).
Both the divergent limiting behavior and our method of interpolation
are described in appendix \ref{app:interpolate}.


\subsection {Results}

Our main numerical results have already been displayed in
figs.\ \ref{fig:lstopPlot} and \ref{fig:sigratioPlot}, which were
generated by using the above formulas for $d\Gamma/d\xi$ in
(\ref{eq:lstopRatioB}) and (\ref{eq:testB}) respectively.
The additional plot fig.\ \ref{fig:dEdzPlot} of the size of NLO
corrections to $dE/dz$ (which is well defined for large-$\Nf$ QED)
is given by
\begin {equation}
   \frac{\Delta(dE/dz)}{(dE/dz)^{(0)}}
   =
   \frac{ \Avg[1-\xi]^{({\rm NLO})} }
        { \Avg[1-\xi]^{(0)} }
   \,,
\end {equation}
since $(1-\xi)E$ is the energy lost when the original electron goes
from energy $E$ to $\xi E$.


\section{Renormalization scale dependence of
         \boldmath$\Delta\lstop/\lstop^{(0)}$}
\label {sec:RenormScale}

We now give a little more detail about the renormalization scale dependence
of our result for $\Delta\lstop/\lstop^{(0)}$.

A physical quantity, like $\lstop$, should not depend on the choice of
renormalization scale $\mu$ if computed to all orders and expressed
in terms of either (i) other physical quantities, or (ii) things that
can in principle be extracted from other physical quantities.
An example of the latter is
MS-bar $\alphaqed(\mu_0)$ at some specific choice of scale $\mu_0$
that is expressed in terms of physical quantities, like
our $\mu_0 \equiv (\hat q E)^{1/4}$ chosen in fig.\ \ref{fig:lstopPlot}.
But generally, when one truncates a small-coupling expansion at some
finite order, the result {\it does} have $\mu$ dependence---the
size of the variation with $\mu$
is formally of order the size of yet-higher-order
corrections which have not been calculated, which in our case would
be NNLO.
But the evaluation of $\lstop$ through NLO order (for constant
$\hat q^{(0)}$) is an exception.  Our expression for
$\lstop^{(0)} + \Delta\lstop$ (but neither term individually) is
actually $\mu$-independent {\it provided} one correspondingly
uses the 1-loop approximation to the renormalization group
equation for $\alpha(\mu)$.

This feature is a consequence of the leading-order result
$\lstop^{(0)}$ being order $\alpha^{-1}$.  Consider the solution
\begin {equation}
   \frac{1}{\alpha(\mu)}
   = \frac{1}{\alpha(\mu_0)} - \beta_0 \ln\Bigl( \frac{\mu}{\mu_0} \Bigr)
\label {eq:alphaRNG1loop}
\end {equation}
to the 1-loop renormalization group equation
\begin {equation}
   \frac{\partial \alpha(\mu)}{\partial \ln\mu}
   = \beta_0 \, \alpha^2(\mu) .
\end {equation}
In order for the stopping length to be $\mu$-invariant through NLO,
and given that $\lstop^{(0)} \propto \alpha^{-1}$, the stopping length
must have an expansion through NLO in $\alpha$ of the form
\begin {equation}
  \lstop \simeq
  \frac{A}{\alpha(\mu)} +
  \left[
    A \beta_0 \ln\Bigl( \frac{\mu}{\mu_0} \Bigr) - B
  \right] ,
\label{eq:lstopAB}
\end {equation}
where $A$ and $B$ are expressions that
depend on $\hat q^{(0)}$ and $E$ but not on
$\mu$ or $\alpha(\mu)$.
But then (\ref{eq:alphaRNG1loop}) means that the right-hand
side of (\ref{eq:lstopAB}) does not depend at all on the value of
$\mu$ if one uses 1-loop renormalization group equations for
$\alpha(\mu)$.

The reason that our results for
\begin {equation}
   \frac{\Delta\lstop}{\lstop^{(0)}}
   =
   \frac{\lstop}{\lstop^{(0)}} - 1
\label {eq:rat}
\end {equation}
nonetheless depend on $\mu$ is because the {\it leading}-order result
$\lstop^{(0)} = A/\alpha(\mu)$ depends on $\mu$ implicitly through
$\alpha(\mu)$.  Note that in fig.\ \ref{fig:lstopPlot}, we have chosen
to keep the physics constant as we vary $\mu$ by taking the horizontal
axis to be $\Nf\alpha(\mu_0)$ rather than $\Nf\alpha(\mu)$.
Specifically, using (\ref{eq:alphaRNG1loop}) and the $\mu$-independence
of (\ref{eq:lstopAB}), we can rewrite
(\ref{eq:rat}) as
\begin {equation}
   \frac{\Delta\lstop}{\lstop^{(0)}}
   =
   -
   \frac{
     (B/A) - \beta_0 \, \ln(\mu/\mu_0)
   }{[\alpha(\mu_0)]^{-1} - \beta_0 \, \ln(\mu/\mu_0) }
   \,.
\label {eq:rat2}
\end {equation}
This expression sheds some light (at least algebraically) on the
special value of $\Nf\alpha(\mu_0)$ in fig.\ \ref{fig:lstopPlot}
where there is no $\mu$ dependence: it corresponds to the value
$\alpha(\mu_0) = A/B = 0.77$, for which the
$\mu$-dependent numerator and denominator cancel each other
in (\ref{eq:rat2}).
[The value of $B$ in (\ref{eq:lstopAB}) given
by our results happens to be positive.]
This exact cancellation of $\mu$ dependence for this special point
is seemingly just an artifact of working to 1-loop order in our
analysis of $\Delta\lstop/\lstop^{(0)}$ and the running of
$\alpha(\mu)$.

There is nothing special about the $\mu=\mu_0$ line in
fig.\ \ref{fig:lstopPlot} being straight for the particular choice
$\mu_0 = (\hat q E)^{1/4}$.  If we had instead plotted $\Nf\,\alpha(2\mu_0)$
on the horizontal axis, it would have been the $\mu=2\mu_0$ line that would
have been straight.

\begin {figure}[t]
\begin {center}
  \includegraphics[scale=1.0]{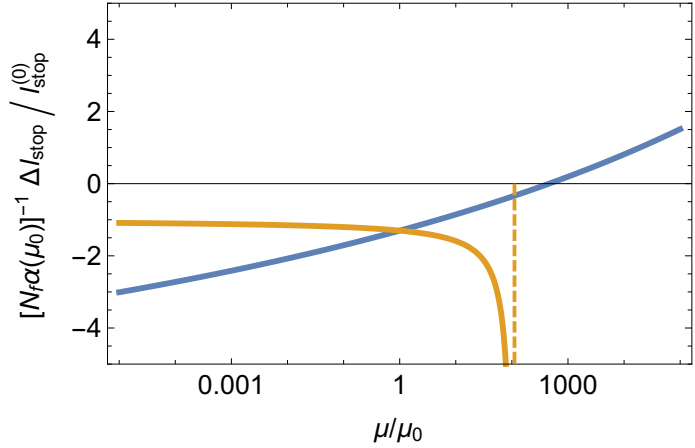}
  \caption{
     \label{fig:lstopPlotMu}
     $\mu$-dependence of $[\Nf\alpha(\mu_0)]^{-1}\,\Delta\lstop/\lstop^{(0)}$
     over a nonsensically
     large range of $\mu$.  The two curves are for
     $\Nf\alpha(\mu_0)=0.1$ (blue) and $\Nf\alpha(\mu_0)=1.0$ (orange).
     The factor of $[\Nf\alpha(\mu_0)]^{-1}$ scaling the vertical
     axis is there just to make the two curves comparable on the same
     plot, given that $\Delta\lstop/\lstop^{(0)}$ is of order $\Nf\alpha$.
  }
\end {center}
\end {figure}

It is interesting to look at how $\Delta\lstop/\lstop^{(0)}$ behaves
over a much wider range of $\mu$ than shown in fig.\ \ref{fig:lstopPlot}.
The $\mu$-dependence is shown in fig.\ \ref{fig:lstopPlotMu} for the two
cases $\Nf\alpha(\mu_0) = 0.1$ and $1.0$, for a range of $\mu$ that gets
nonsensically far away from $\mu_0$.  The vertical dashed line denotes
the location,
in the case $\Nf\alpha(\mu_0)=1.0$, where
the 1-loop $\alpha(\mu)$ blows up to $+\infty$ and continuum
QED breaks down as
a sensible theory:
the Landau pole
\begin {equation}
   \Lambda_{\rm L} = \mu_0 \, e^{1/\beta_0\alpha(\mu_0)} .
\end {equation}
Note here that even for $\Nf\alpha(\mu_0) = 1.0$, the
Landau pole is still two orders of magnitude away from the
physics scale $\mu_0$, and so our calculations are sensible.
Another way of saying this is that for some sorts of physics,
like the running of the coupling, $\Nf\alpha(\mu_0) = 1.0$ can be
considered a somewhat small coupling,
whereas for the question of whether or not overlapping
formation time effects are negligible in a medium, we have seen from
figs.\ \ref{fig:lstopPlot} and \ref{fig:sigratioPlot} that
it is a moderately large coupling.

\section{Caveats for \boldmath$\test$ test of strong vs.\ weak-coupled
         splitting}
\label{sec:caveats}

In the introduction, we motivated our discussion of the
relative NLO correction $\test$ in the expansion
$\sigma/\lstop \simeq (\sigma/\lstop)^{(0)} (1 + \test)$ as
a (partly) successful method for testing the size of NLO corrections
that could not be absorbed into $\hat q$.
Here we explain the caveat ``partly,'' which applies to
the situation of double-log corrections to $\hat q$ in QCD.

The idea was that corrections to $\hat q$ would cancel between
the numerator and denominator in $\sigma/\lstop$.  This is true
of energy-independent corrections to $\hat q$, but the situation
is more subtle for corrections that depend logarithmically on
energy.


\subsection{Warm up / review for a single logarithm}

It will be instructive to imagine a theory where the
NLO corrections to $\hat q$ depended on a {\it single} logarithm
of energy that, so that
\begin {subequations}
\label {eq:1log1}
\begin {equation}
   \delta\hat q(E) \approx A \alpha \hat q^{(0)} \ln(E/\hat q\tau_0^2)
   \qquad
   \mbox{(for illustration only)}
\end {equation}
for some $O(1)$ constant $A$,
giving
\begin {equation}
   \frac{d\Gamma^{\rm(NLO)}}{d\xi}(E,\xi) \approx
   \frac{d\Gamma^{(0)}}{d\xi}(E,\xi) \times
   \tfrac12 A \ln(E/\hat q\tau_0^2)
   \qquad
   \mbox{(for illustration only)}
\label {eq:1log1dG}
\end {equation}
\end {subequations}
since $d\Gamma^{(0)}/d\xi$ scales like $\sqrt{\hat q/E}$.
In this paper, depending on context, the symbol $E$ sometimes
refers (as above) to the energy of the parent of some splitting or overlapping
double-splitting somewhere in the shower, and it sometimes instead
refers to the energy of the particle that initiated the entire shower.
To avoid confusion, let's call the latter $E_0$ here, and write
$E = y E_0$ for the parent of the NLO splitting in (\ref{eq:1log1dG}),
where $y$ is the longitudinal
momentum fraction of that splitting's parent relative to
the total energy of the shower.
Then (\ref{eq:1log1}) may be rewritten as
\begin {subequations}
\begin {align}
   \delta\hat q(E) &\approx A \alpha \hat q^{(0)} \ln(y E_0/\hat q\tau_0^2)
\nonumber\\
   &= A \alpha \hat q^{(0)} \ln(E_0/\hat q\tau_0^2)
      +
      A \alpha \hat q^{(0)} \ln y
\nonumber\\
   &= (\mbox{large logarithm constant})
      +
      A \alpha \hat q^{(0)} \ln y ,
\label {eq:1log2qhat}
\end {align}
giving
\begin {equation}
   \frac{d\Gamma^{\rm(NLO)}}{d\xi}(E,\xi) \approx
   \frac{d\Gamma^{(0)}}{d\xi}(E,\xi) \times
   (\mbox{large logarithm constant})
   +
   \frac{d\Gamma^{(0)}}{d\xi}(E,\xi) \times
   \tfrac12 A \ln y .
\label {eq:1log2dG}
\end {equation}
\end {subequations}
The potentially large first term in (\ref{eq:1log2qhat}) is
a constant addition to $\hat q$ and so the first term in
(\ref{eq:1log2dG}) should cancel in the calculation of the
ratio $\sigma/\lstop$.  We've seen that explicitly in the discussion
after (\ref{eq:testA}).
The second term in (\ref{eq:1log2dG}) does not generate large
logarithms because, as discussed in the introduction with regard to
(\ref{eq:series}),
the late-time evolution of the shower
(small $y$ above) does not contribute significantly to stopping
distances.  In formulas, the effect of the $\ln y$ term in
(\ref{eq:1log2dG}) would be precisely the
$\Avg[\cdots \ln\xi]^{(0)}$ terms in (\ref{eq:testB})
[with the constant ``$\frac14\beta_0$'' there identified as the
$\frac12 A$ of (\ref{eq:1log1dG})].
Those $\Avg[\cdots \ln\xi]^{(0)}$ terms represent convergent
integrals over $\xi$ and do
not generate any logarithmic enhancement of the result for $\test$.


\subsection{Double logarithms}

Now consider in contrast a double-log correction to $\hat q$, as in QCD:
\begin {align}
   \delta\hat q(E) &\approx A \alpha \hat q^{(0)} \ln^2(y E_0/\hat q\tau_0^2)
\nonumber\\
   &= A \alpha \hat q^{(0)} \ln^2(E_0/\hat q\tau_0^2)
      +
      2 A \alpha \hat q^{(0)} \ln(E_0/\hat q\tau_0^2) \, \ln y
      +
      A \alpha \hat q^{(0)} \ln^2 y
\nonumber\\
   &= (\mbox{double-log enhanced constant})
      +
      (\mbox{single-log enhanced constant}) \ln y
\nonumber\\ & \qquad
      +
      A \alpha \hat q^{(0)} \ln^2(y) .
\label {eq:2log2qhat}
\end {align}
The first term above, which is the largest, is a constant and so will
not contribute to $\sigma/\lstop$ and so will not affect $\test$.
The last term will not give any large-log contribution to
$\sigma/\lstop$ because stopping distances are not sensitive to
small $y$.  But the middle term can be expected to give a
{\it single}-log enhanced contribution to $\sigma/\lstop$ and so
to $\test$.

We characterized our proposal of studying $\test$ as only partly
successful because it manages to avoid double-log
corrections (as well as avoiding all non-log corrections to $\hat q$),
but there will be an issue with sub-leading logarithms in QCD.
For QED, there are no double logs, and so this difficulty does
not arise there.


\section{Conclusion}
\label {sec:conclusion}

In this paper, we have proposed various simple characteristics
of in-medium showers that can be used to test the size of overlapping
formation time corrections, in an attempt to distinguish
weakly vs. strongly-coupled splitting pictures of shower
development.
Part of our motivation was to find a measure insensitive
to corrections that can be absorbed into $\hat q$.
We found one for QED, but our proposal seems like it would
run into difficulty at the level of sub-leading logarithms
in QCD.  This underscores the importance of understanding how
to account for and absorb subleading logarithms in ongoing
research on the calculation and structure of overlapping
formation time corrections in QCD.

We used large-$\Nf$ QED as a concrete example in this paper,
finding, for example, that the effect of overlapping
formation times is modest
(the weak-splitting picture appears good) for
$\Nf\alphaqed(Q_\perp) \simeq 0.2$, but is roughly a 100\% effect for
$\Nf\alphaqed(Q_\perp) \simeq 1$.
One is of course tempted to rashly speculate about how this might
or might not translate to the
QCD analog $\CA\alphas {=} \Nc\alphas$ of
$\Nf\alphaqed$, but that would be premature.
For one thing, the infrared behavior of the LPM effect in QED
and QCD is very different.
The completed set of rates needed for a (large $\Nc$)
QCD NLO calculation of shower characteristics will hopefully
be available in the future, but the issue of subleading logarithms
will then need to be understood.

In large-$\Nf$ QED, the results of fig.\ \ref{fig:lstopPlot}
indicate that the net effect of overlapping formation times
is to {\it reduce} the stopping length
(corresponding to an increase in $dE/dz$).
It would be nice to have a simple physical picture for understanding
this sign of the result.


\begin{acknowledgments}

We thank Yacine Mehtar-Tani for useful conversations
and the CERN theory group for their hospitality during the time
this work was completed.
This work was supported, in part, by the U.S. Department
of Energy under Grant No.~DE-SC0007984.

\end{acknowledgments}

\appendix

\section{Energy stopping distance and other generalizations}
\label {app:energy}

Here, we will generalize the analysis of section \ref{sec:lstopNoLog}
to moments $\langle z^n\rangle$
of the {\it energy} deposition distribution $\varepsilon(z)$, though
we will not follow through with numerics in this paper.
We will also then adapt the analysis to discuss formulas for
moments of the charge deposition distribution when {\it not}\/ in a
large-$N$ limit (e.g.\ for $\Nf{=}1$ QED).


\subsection {Energy deposition}

In what follows, we use the term ``species'' to distinguish
different types of particles: e.g.\ quark vs.\ gluon in QCD or
electron vs.\ photon in QED.


\subsubsection {$1{\to}2$ splitting: single species}

We start by imaging the simplest case: only $1{\to}2$ splittings
of particles in the shower, like in leading-order calculations of
shower development.  We also start by considering just one species
of particle, like a shower initiated by a gluon in large-$\Nc$ QCD,
where only $g \to gg$ would be relevant.  The analog, for energy
deposition, of
the charge deposition equation
(\ref{eq:rhoeq0}) is
\begin {equation}
  \frac{\partial \varepsilon(E,z)}{\partial z}
  =
  - \Gamma(E) \, \varepsilon(E,z)
  +
  \frac12
  \int_0^1 d\xi \> \frac{d\Gamma}{d\xi}(E,\xi) \,
  \Bigl[
    \xi\,\varepsilon(\xi E,z)
     + (1{-}\xi)\,\varepsilon\bigl((1{-}\xi)E,z\bigr)
  \Bigr] 
  .
\label {eq:epseq1}
\end {equation}
Here $\xi$ and $1{-}\xi$ represent the longitudinal momentum fractions
of the two daughters in a splitting.  The overall factor of $\frac12$
in front of the integral is the final-state identical-particle
symmetry factor, since we are focusing for the moment on the case
that the two daughters (as well as the parent) of any splitting
are the same species.  There are two terms in the integrand, as opposed
to just one in (\ref{eq:rhoeq0}), because {\it both} daughters carry
some of the energy of the parent, so the future evolution of both
contribute to where the energy of the shower will be deposited.
Since the energies of the two daughters are weighted by
$\xi$ and $1{-}\xi$ respectively, their individual distributions of
where they deposit their energy should be weighted accordingly in
their contribution to where the parent distributes its energy,
which is the origin of the factors $\xi$ and $1{-}\xi$ multiplying
the corresponding daughter's $\varepsilon$ inside the square brackets
above.

One may combine terms to rewrite (\ref{eq:epseq1}) as
\begin {equation}
  \frac{\partial \varepsilon(E,z)}{\partial z}
  =
  -
  \frac12
  \int_0^1 d\xi \> \frac{d\Gamma}{d\xi}(E,\xi) \,
  \Bigl[
    \varepsilon(E,z)
    - \xi\,\varepsilon(\xi E,z)
    - (1{-}\xi)\,\varepsilon\bigl((1{-}\xi)E,z\bigr)
  \Bigr] 
  .
\label {eq:epseq1b}
\end {equation}
One could also use the $\xi \leftrightarrow 1{-}\xi$ symmetry
associated with having two identical daughters to simplify to
\begin {equation}
  \frac{\partial \varepsilon(E,z)}{\partial z}
  =
  -
  \frac12
  \int_0^1 d\xi \> \frac{d\Gamma}{d\xi}(E,\xi) \,
  \Bigl[
    \varepsilon(E,z)
    - 2 \xi\,\varepsilon(\xi E,z)
  \Bigr] 
  ,
\end {equation}
but, for the sake of later generalization, we will find it useful to
stay with (\ref{eq:epseq1b}).

Now consider the case where rates may be taken to scale with energy
as $E^{-\nu}$, as in sections \ref{sec:lstopNoLog} and \ref{sec:lstop1Log}
of the main text.  Write
\begin {equation}
  \varepsilon(E,z) = E^{-\nu} \, \tilde\varepsilon(E^{-\nu}z)
\end {equation}
analogous to (\ref{eq:rhoscale}).  This yields
\begin {equation}
  \frac{d\tilde\varepsilon(\tilde z)}{d\tilde z}
  =
  -
  \frac12
  \int_0^1 d\xi \> \frac{d\tilde\Gamma}{d\xi} \,
  \Bigl[
    \tilde\varepsilon(\tilde z)
    - \xi^{1-\nu}\,\tilde\varepsilon(\xi^{-\nu}\tilde z)
    - (1{-}\xi)^{1-\nu}\,\tilde\varepsilon\bigl((1{-}\xi)^{-\nu}\tilde z\bigr)
  \Bigr] 
\end {equation}
as the energy-deposition analog of (\ref{eq:rhoeqscaled}).
Taking moments of both sides leads to the recursion relation
\begin {equation}
   \langle \tilde z^n \rangle =
   \frac{ n \langle \tilde z^{\,n-1} \rangle }
        {
          \frac12
          \int_0^1 d\xi \> \frac{d\tilde\Gamma}{d\xi}
          [1-\xi^{1+n\nu}-(1{-}\xi)^{1+n\nu}]
        }
   \,.
\end {equation}
Compare to (\ref{eq:recursionA}).
Note that in this appendix
(except for section \ref{app:chargeNf1}) the notation
$\langle z^n \rangle$ refers to moments
of the energy deposition distribution $\varepsilon$, whereas
elsewhere in this paper it refers to
moments of the charge distribution $\rho$.

One could now expand in $\alpha$ to obtain formulas for NLO
corrections similar to
(\ref{eq:lstopRatioA}), (\ref{eq:testA}), (\ref{eq:lstopRatioB}), and
(\ref{eq:testB}), but we will not go into that level of detail
here.


\subsubsection {$1{\to}2$ splitting: multiple species}

Now allow for multiple species, such as quarks and gluons.
Showers initiated by different species $i$ will generally produce
different distributions $\varepsilon_i(E,z)$.
The corresponding generalization of (\ref{eq:epseq1}) and
(\ref{eq:epseq1b}) is
\begin {align}
  \frac{\partial \varepsilon_i(E,z)}{\partial z}
  &=
  - \Gamma_i(E) \, \varepsilon_i(E,z)
  +
  \frac12 \sum_{jk}
  \int_0^1 d\xi \>
  \frac{d\Gamma_{i\to jk}}{d\xi}
     \bigl(E{\to}\xi E, (1{-}\xi)E\bigr) \,
\nonumber\\ & \hspace{10em} \times
  \Bigl[
    \xi\,\varepsilon_j(\xi E,z)
     + (1{-}\xi)\,\varepsilon_k\bigl((1{-}\xi)E,z\bigr)
  \Bigr]
\nonumber\\
  &=
  -
  \frac12 \sum_{jk}
  \int_0^1 d\xi \>
  \frac{d\Gamma_{i\to jk}}{d\xi}
     \bigl(E{\to}\xi E, (1{-}\xi)E\bigr) \,
\nonumber\\ & \hspace{10em} \times
  \Bigl[
     \varepsilon_i(E,z)
     - \xi\,\varepsilon_j(\xi E,z)
     - (1{-}\xi)\,\varepsilon_k\bigl((1{-}\xi)E,z\bigr)
  \Bigr]
  .
\label {eq:epseq2}
\end {align}
Here, for $j{=}k$ terms of the double sum over species
(e.g. $g{\to}gg$ in QCD),
the overall factor of $\frac12$
in front of the integral is again a final-state identical-particle
symmetry factor.  For the $j{\not=}k$ terms of the sum, the
overall factor of $\frac12$ is canceled by the two permutations
of $jk$ in the double sum.  For example, for $i$ an electron in
QED, the sum over $j$ and $k$ on the right-hand side
would contain two terms which would be exactly equal:
$e{\to}\gamma e$ and $e{\to}e \gamma$.
The summation in (\ref{eq:epseq2}) is just a compact way of
simultaneously accounting for the cases of identical and
non-identical daughters.

Scaling out the energy dependence and taking moments of the equation,
one finds
\begin {align}
  -n \langle\tilde z^{n-1}\rangle_i
  &=
  -
  \frac12 \sum_{jk}
  \int_0^1 d\xi \> \frac{d\tilde\Gamma_{i\to jk}}{d\xi}
  \bigl[ \langle \tilde z^n \rangle_i
         - \xi^{1+n\nu} \langle \tilde z^n \rangle_j
         - (1{-}\xi)^{1+n\nu} \langle \tilde z^n \rangle_k
  \bigr]
\nonumber\\
  &\equiv
  - \sum_m (M_{(n)})_{im} \langle \tilde z^n \rangle_m
\end {align}
As indicated by the last line, it is convenient to write the
relation in terms of a matrix $M_{(n)}$ acting on species space,
with matrix elements
\begin {equation}
   (M_{(n)})_{i m} =
  \frac12 \sum_{jk}
  \int_0^1 d\xi \> \frac{d\tilde\Gamma_{i\to jk}}{d\xi}
  \bigl[ \delta_{im}
         - \xi^{1+n\nu} \delta_{jm}
         - (1{-}\xi)^{1+n\nu} \delta_{km}
  \bigr] .
\end {equation}
The recursion relation for the moments is then given in terms of the
matrix inverse:
\begin {equation}
   \langle \tilde z^n \rangle_i =
   n \sum_j (M_{(n)}^{-1})_{ij} \langle \tilde z^{n-1} \rangle_j
   \,.
\label {eq:recursionM}
\end {equation}
The $n{=}1$ case reproduces (for $\nu=\frac12$) the energy stopping
length formula derived in ref.\ \cite{stop} for the
case of quarks and gluons.


\subsubsection {$1{\to}2$ and $1{\to}3$ splitting}

The point of this paper is to be able to account for next-to-leading
order corrections to shower development, which are generated when two
consecutive splittings overlap.  As discussed in section I.A of
ref.\ \cite{seq}, overlap corrections to double splitting should be
treated as effectively a type of $1{\to}3$ splitting in order to
analyze shower development in the framework of classical probability
theory.%
\footnote{
  Depending on whether overlap effects reduce or enhance the
  double-splitting process,
  the ``rate'' assigned to this effective $1{\to}3$ splitting
  process could be negative.
  But that does not cause any difficulty for the type of
  analysis considered in this paper.
}
In addition, there are also direct $1{\to}3$ splitting processes
at NLO due, for example, to the 4-gluon vertex in QCD \cite{4point}
or interactions involving intermediate longitudinally-polarized
gauge bosons \cite{QEDnf}.

Let $d\Gamma_{i\to jk}$ and $d\Gamma_{i\to jkl}$ be the rates for
$1{\to}2$ and effective $1{\to}3$ processes respectively.
The generalization of (\ref{eq:epseq2}) to include the
$1{\to}3$ processes is
\begin {align}
  \frac{\partial \varepsilon_i(E,z)}{\partial z}
  =
  &-
  \frac12 \sum_{jk}
  \int_0^1 d\xi_1 \> d\xi_2 \> \delta(1{-}\xi_1{-}\xi_2) \>
  \frac{d\Gamma_{i\to jk}}{d\xi_1}
     \bigl(E{\to}\xi_1 E, \xi_2 E\bigr) \,
\nonumber\\ & \hspace{7em} \times
  \Bigl[
     \varepsilon_i(E,z)
     - \xi_1\,\varepsilon_j(\xi_1 E,z)
     - \xi_2\,\varepsilon_k(\xi_2 E,z)
  \Bigr]
\nonumber\\
  &-
  \frac1{3!} \sum_{jkl}
  \int_0^1 d\xi_1 \> d\xi_2 \> d\xi_3 \> \delta(1{-}\xi_1{-}\xi_2{-}\xi_3) \>
  \frac{d\Gamma_{i\to jkl}}{d\xi_1\,d\xi_2}
     \bigl(E{\to}\xi_1 E, \xi_2 E, \xi_3 E\bigr) \,
\nonumber\\ & \hspace{7em} \times
  \Bigl[
     \varepsilon_i(E,z)
     - \xi_1\,\varepsilon_j(\xi_1 E,z)
     - \xi_2\,\varepsilon_k(\xi_2 E,z)
     - \xi_3\,\varepsilon_l(\xi_3 E,z)
  \Bigr]
  ,
\label {eq:epseq3}
\end {align}
where we have now used $\delta$-functions to write the integrals over
daughter momentum fractions $\xi$ in a more symmetric form.
For rates that scale with energy as $E^{-\nu}$, the recursion relation
for the moments is again (\ref{eq:recursionM}) but now with
\begin {equation}
   (M_{(n)})_{i m} =
   \Avg_{1\to2}[
     \delta_{im}
     - \xi_1^{1+n\nu} \delta_{jm}
     - \xi_2^{1+n\nu} \delta_{km}
   ]
   +
   \Avg_{1\to 3}[
     \delta_{im}
     - \xi_1^{1+n\nu} \delta_{jm}
     - \xi_2^{1+n\nu} \delta_{km}
     - \xi_3^{1+n\nu} \delta_{lm}
   ]
\label {eq:Mresult3}
\end {equation}
with
\begin {equation}
  \Avg_{1\to2}[f_{i{\to}jk}(\xi_1,\xi_2)] \equiv
  \frac12 \sum_{jk}
  \int_0^1 d\xi_1 \> d\xi_2 \> \delta(1{-}\xi_1{-}\xi_2) \>
  \frac{d\tilde\Gamma_{i\to jk}}{d\xi_1}
     \bigl(1{\to}\xi_1, \xi_2) \,
  f_{i{\to}jk}(\xi_1,\xi_2) ,
\end {equation}
\begin {multline}
  \Avg_{1\to3}[f_{i{\to}jkl}(\xi_1,\xi_2,\xi_3)] \equiv
  \frac1{3!} \sum_{jkl}
  \int_0^1 d\xi_1 \> d\xi_2 \> d\xi_3 \> \delta(1{-}\xi_1{-}\xi_2{-}\xi_3) \>
  \frac{d\tilde\Gamma_{i\to jkl}}{d\xi_1\,d\xi_2}
     \bigl(1{\to}\xi_1, \xi_2, \xi_3\bigr) \,
\\ \times
  f_{i{\to}jkl}(\xi_1,\xi_2,\xi_3) .
\label {eq:Avg3}
\end {multline}


\subsubsection {Cancellation of QCD power-law divergences}

In the above formulas, we have (i) grouped overlap corrections to
actual double-splitting (e.g.\ fig.\ \ref{fig:diags}) into
$d\Gamma_{i\to jkl}$ and (ii) grouped the corresponding NLO virtual corrections
to single splitting (e.g.\ fig.\ \ref{fig:diagsVIRT}) together
with leading-order single splitting into $d\Gamma_{i\to jk}$.
However, in QCD, these LPM rates individually have severe
power-law (not just logarithmic) infrared divergences that
make the two terms in (\ref{eq:Mresult3}) separately infinite.
Specifically, refs.\ \cite{2brem,seq} show for
large-$\Nc$ QCD that%
\footnote{
  Our $\xi_2$ and $\xi_3$ here are the $x$ and $y$ of
  refs.\ \cite{2brem,seq}.
}
\begin {equation}
   \frac{d\tilde\Gamma_{g\to ggg}}{d\xi_2\,d\xi_3}
   \sim
   \frac{\Nc^2\alphas^2}{\xi_2 \xi_3^{3/2}}
   \sqrt{\frac{\hat q}{E}}
   \qquad
   \mbox{for $\xi_3 \ll \xi_2 \ll 1$.}
\label {eq:blowup}
\end {equation}
(See section I.D of ref.\ \cite{seq} for a qualitative discussion.)
As noted there, this blows up fast enough as $\xi_3 \to 0$
to have a divergent effect on energy loss if NLO virtual corrections
to single splitting are ignored.  In particular, (\ref{eq:blowup})
means that the $\Avg_{1\to3}[\cdots]$ in (\ref{eq:Mresult3}) has
a power-law divergence from, for example,
the $\xi_3 \to 0$ region of integration in (\ref{eq:Avg3}).
It is firmly expected that this QCD power-law divergence will cancel
in the sum in (\ref{eq:Mresult3}) of $\Avg_{1\to 3}[\cdots]$
and the NLO piece of $\Avg_{1\to 2}[\cdots]$, leaving only
the known QCD double-log divergence.
This is something to verify in the future when
results are available for the NLO virtual corrections to
QCD single splitting.  (The double log term itself needs to be treated using
a more sophisticated method for computing the moments
$\langle z^n \rangle$, as discussed in section \ref{sec:dbllog}.)


\subsection {Charge deposition without assuming large \boldmath$N$}
\label {app:chargeNf1}

Using the same methods as above, we can also generalize the charge
deposition discussion
of section \ref{sec:lstopNoLog} to handle the case where $\Nf$ is not
large.  Beyond leading-order,
we must then follow all of the charged daughters of each
splitting because of interference effects like
fig.\ \ref{fig:ambiguity}a.

To adapt (\ref{eq:epseq3}) to
charge deposition, it's useful to normalize each $\rho_i(E,z)$
so that its integral (i.e.\ $\langle 1 \rangle_i$)
gives the relevant charge of species $i$.  Then the charge
deposition analog
of (\ref{eq:epseq3}) is 
\begin {align}
  \frac{\partial \rho_i(E,z)}{\partial z}
  =
  &-
  \frac12 \sum_{jk}
  \int_0^1 d\xi_1 \> d\xi_2 \> \delta(1{-}\xi_1{-}\xi_2) \>
  \frac{d\Gamma_{i\to jk}}{d\xi_1}
     \bigl(E{\to}\xi_1 E, \xi_2 E\bigr) \,
\nonumber\\ & \hspace{7em} \times
  \Bigl[
     \rho_i(E,z)
     - \rho_j(\xi_1 E,z)
     - \rho_k(\xi_2 E,z)
  \Bigr]
\nonumber\\
  &-
  \frac1{3!} \sum_{jkl}
  \int_0^1 d\xi_1 \> d\xi_2 \> d\xi_3 \> \delta(1{-}\xi_1{-}\xi_2{-}\xi_3) \>
  \frac{d\Gamma_{i\to jkl}}{d\xi_1\,d\xi_2}
     \bigl(E{\to}\xi_1 E, \xi_2 E, \xi_3 E\bigr) \,
\nonumber\\ & \hspace{7em} \times
  \Bigl[
     \rho_i(E,z)
     - \rho_j(\xi_1 E,z)
     - \rho_k(\xi_2 E,z)
     - \rho_l(\xi_3 E,z)
  \Bigr]
  .
\label {eq:rhoeqNf1}
\end {align}
However, in QED (and analogously in QCD), charge conjugation invariance
gives
\begin {equation}
  \rho_\gamma(E,z) = 0 ,
  \qquad
  \rho_{\bar e}(E,z) = -\rho_e(E,z) ,
\end {equation}
and so one may rewrite (\ref{eq:rhoeqNf1}) as an equation just for
$\rho_e(E,z)$:
\begin {align}
  \frac{\partial \rho_e(E,z)}{\partial z}
  =
  &-
  \int_0^1 d\xi_\gamma \> d\xi_e \> \delta(1{-}\xi_\gamma{-}\xi_e) \>
  \frac{d\Gamma_{e\to \gamma e}}{d\xi_{e}} \,
\nonumber\\ & \hspace{7em} \times
  \Bigl[
     \rho_e(E,z)
     - \rho_e(\xi_e E,z)
  \Bigr]
\nonumber\\
  &-
  \frac12
  \int_0^1 d\xi_{e1} \> d\xi_{e2} \> d\xi_{\bar e} \>
  \delta(1{-}\xi_{e1}{-}\xi_{e2}{-}\xi_{\bar e}) \>
  \frac{d\Gamma_{e\to ee\bar e}}{d\xi_{e1}\,d\xi_{e2}} \,
\nonumber\\ & \hspace{7em} \times
  \Bigl[
     \rho_e(E,z)
     - \rho_e(\xi_{e1} E,z)
     - \rho_e(\xi_{e2} E,z)
     + \rho_e(\xi_{\bar e} E,z)
  \Bigr]
  .
\end {align}
For rates that scale as $E^{-\nu}$, the corresponding recursion relation
for the moments of $\rho_e$ would be
\begin {equation}
   \langle \tilde z^n \rangle =
   \frac{ n \langle \tilde z^{n-1} \rangle }
        {
          \Avg_{e\to\gamma e}[ 1 - \xi_e^{n\nu} ]
          +
          \Avg_{e\to ee\bar e}
            [ 1 - \xi_{e1}^{n\nu} - \xi_{e2}^{n\nu} + \xi_{\bar e}^{n\nu} ]
         }
   \,.
\end {equation}


\section{Numerical solution for leading-order \boldmath$\rho^{(0)}(z)$}
\label {app:rhoNumeric}

In this appendix, we discuss how to numerically solve for the
full {\it leading}-order charge-stopping distribution $\rho^{(0)}(z)$,
which is interesting but not necessary for anything else in this
paper.  We will not have to restrict attention to the large-$\Nf$
limit of QED in this appendix: large $\Nf$ was used to simplify
the complexity of NLO calculations but this appendix only deals with
leading order.


\subsection {Numerical Procedure}

In particular, we will solve the scaled equation (\ref{eq:rhoeqscaled}),
which is directly relevant only to leading-order calculations (given
our assumptions) but not to NLO (which contains log dependence on
energy).  Note that (\ref{eq:rhoeqscaled}) is linear in $\tilde\rho$, and so
solving that equation does not by itself determine the overall
normalization of $\tilde\rho$.  But we may numerically find a solution with
${\it any}$ normalization and then, as a final step, compute the
normalization ${\cal N} = \int_0^\infty d\tilde z\>\tilde\rho(\tilde z)$
and then rescale $\tilde\rho(z)$ by a factor of $1/{\cal N}$ to make
$\tilde\rho$ into a normalized probability distribution.

To solve numerically, discretize $z$ and rewrite
(\ref{eq:rhoeqscaled}) as
\begin {equation}
  \tilde\rho(\tilde z{-}\Delta\tilde z)
  =
  \tilde\rho(\tilde z)
  +
  \Delta\tilde z 
  \int_0^1 d\xi \> \frac{d\tilde\Gamma}{d\xi} \,
  \bigl[ \tilde\rho(\tilde z)
         - \xi^{-1/2} \tilde\rho\bigl(\xi^{-1/2}\tilde z\bigr) \bigr]
\label {eq:rhoeqscaledN}
\end {equation}
for small steps $-\Delta\tilde z$ and $\nu=\frac12$.
This version of the equation
lets us take a solution for large $\tilde z$ (say $\tilde z >$ some $Z$)
and extrapolate it step
by step to smaller $\tilde z$.
All we need to start the process is an approximate (un-normalized)
solution to (\ref{eq:rhoeqscaled}) in the large-$z$ limit.


\subsection {Large-\boldmath$\tilde z$ solution}

We expect the charge deposition distribution should decay rapidly
(e.g.\ exponentially) for large $z$, so let us make a WKB-inspired
rewriting
\begin {equation}
   \tilde\rho^{(0)}(\tilde z) \equiv e^{-W(\tilde z)} ,
\label {eq:Wdef}
\end {equation}
where we will treat the exponent $W(\tilde z)$ as large.
Plugging (\ref{eq:Wdef}) into (\ref{eq:rhoeqscaled}),
\begin {equation}
  W'(\tilde z) =
  \int_0^1 d\xi \> \frac{d\tilde\Gamma^{(0)}}{d\xi} \,
  \Bigl[ 1 - \xi^{-1/2} e^{W(\tilde z)-W(\xi^{-1/2}\tilde z)} \Bigr].
\label {eq:Weq1}
\end {equation}  
We've explicitly written the superscripts ``$(0)$'' at this point to remind
us that we are only solving the equation for the leading-order result.
We've made that reminder because it will be important here that
the LPM suppression of leading-order
soft bremsstrahlung rates is very different in
QED and QCD, such that the total LPM bremsstrahlung
rate $\Gamma^{(0)}$ is infrared divergent in QCD but not QED.
For finite total $\Gamma^{(0)}$, we assert (and will justify
{\it a posteriori}) that
the second term in the integrand of (\ref{eq:Weq1}) can be ignored
when $\tilde z$ is large, leaving
\begin {equation}
  W'(\tilde z) \simeq
  \int_0^1 d\xi \> \frac{d\tilde\Gamma^{(0)}}{d\xi} \,
  = \tilde\Gamma^{(0)}
\end {equation}  
and so
\begin {equation}
  W(\tilde z) \simeq \tilde\Gamma^{(0)} \tilde z
\end {equation}
and
\begin {equation}
  \tilde\rho^{(0)}(\tilde z) \simeq e^{-\tilde\Gamma^{(0)} \tilde z}
  \qquad
  \mbox{(for large $\tilde z$)} .
\label {eq:rhoasymp}
\end {equation}
Putting this solution back into (\ref{eq:Weq1}), one finds that,
for $\tilde z \gg 1/\tilde\Gamma_{(0)}$,
the second term in the integrand is indeed ignorable compared to
the first unless $1{-}\xi \lesssim (\tilde\Gamma^{(0)}\tilde z)^{-1}$.
That leaves only a parametrically small portion of the $\xi$
integration where the second term is important.  As long as
the integrals of the first and second terms are separately
convergent, that means that the effect of the second term on
the integral is negligible in the large $\tilde z$ limit.

For leading-order QCD, one needs a different analysis, which we
will not present here.


We have not bothered to go to higher order in our WKB-like expansion to
determine whether there is any $\tilde z$ dependence to the
{\it pre-factor} of the exponential behavior (\ref{eq:rhoasymp}).
For large enough $\tilde z$, an algebraic pre-factor will not vary
significantly over the range $1/\tilde\Gamma^{(0)}$ it takes for the
exponential to drop drastically.  So our numerics will not be
very sensitive to that pre-factor provided we only use (\ref{eq:rhoasymp})
for large enough $\tilde z$.  This insensitivity may be checked by varying
the cut-off $Z$ chosen to switch between the asymptotic form
for $\tilde z > Z$ and the numerical evolution using
(\ref{eq:rhoeqscaledN}) for $\tilde z < Z$.


\subsection {Result}

\begin {figure}[t]
\begin {center}
  \begin{picture}(290,195)(0,0)
  \put(17,10){\includegraphics[scale=0.4]{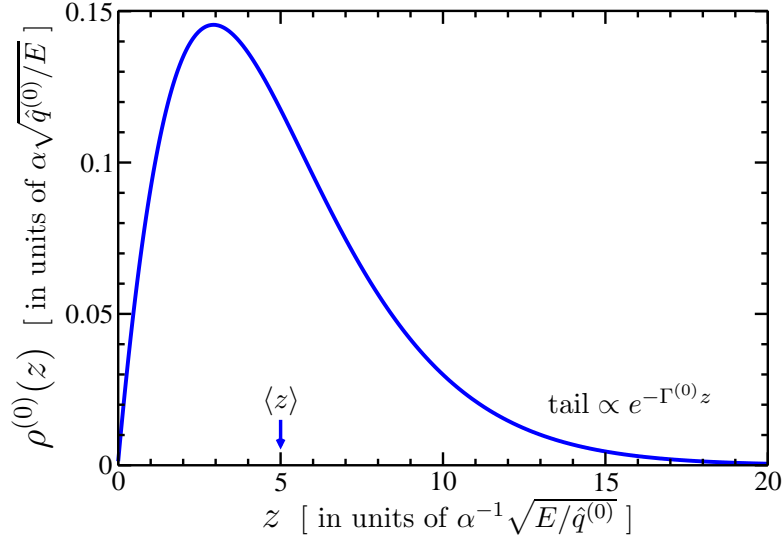}}
  \put(92,42){$\langle z \rangle$}
  \put(88,-3){
      {\large$z$}~~[ in units of $\alpha^{-1} \sqrt{E/\hat q^{(0)}}$ ]
  }
  \put(-3,23){\rotatebox{90}{
      {\large$\rho^{(0)}(z)$}~~[ in units of $\alpha\sqrt{\hat q^{(0)}/E}$ ]
  }}
  \put(200,40){$\mbox{tail} \propto e^{-\Gamma^{(0)} z}$}
  \end{picture}
  \caption{
     \label{fig:deposit2}
     The leading-order result for the distribution of charge deposition
     in large-$\Nf$ QED.  This calculation is made
     in the approximation of constant $\hat q^{(0)}$, as discussed in
     section \ref{sec:MoreAssumptions}.
     The first moment and width of the above curve is
     $\lstop = 5.01$ and $\sigma = 3.24$
     [in units of $\alpha^{-1}(E/\hat q^{(0)})^{1/2}$],
     in agreement with results calculated using the recursion relation
     (\ref{eq:recursionA}) with $d\Gamma^{(0)}/d\xi$.
  }
\end {center}
\end {figure}

Using the above procedure with the QED leading-order LPM bremsstrahlung
rate (\ref{eq:LOformula}) gave the particular
curve shown in fig.\ \ref{fig:deposit}.%
\footnote{
  We found that we converged to reasonably precise, stable results using
  $Z = 20$ and $\Delta z = 0.02$.
}
Since this is a precise numeric result, we now show it in a little more
detail in fig.\ \ref{fig:deposit2}.

\section{One method for charge stopping with double logarithms, etc.}
\label {app:dbllog}

\subsection {Basic equations}

Instead of directly writing an equation (\ref{eq:rhoeq}) for the
charge deposition, consider following the original charge through time,
even before it is deposited, from energy $E$ initially to some
smaller $xE$ at time $t=z$.  Let $P(x,E,t)$ be the probability distribution
for $x$ at time $t$.
We can find an equation for $P(x,E,t)$ somewhat similar to how we
found the equation (\ref{eq:rhoeq}) for $\rho(x,E)$, and very similar
to the method used by ref.\ \cite{Nevolve1} for gluon distributions.
The basic equation for the evolution of $P$ over an infinitesimal
time interval $\Delta t$ is
\begin {equation}
   P(x,E,t{+}\Delta t)
   =
   \bigl[1 - \Gamma(xE)\,\Delta t\bigr] \, P(x,E,t)
   +
   \int_{x}^1 dy \int_0^1 d\xi \,
   \frac{d\Gamma}{d\xi}(\xi, yE) \, \Delta t
   \, P(y,E,t) \, \delta(x-\xi y) .
\end {equation}
The first term is the probability that nothing happened in the
small time interval $[t,t{+}\Delta t]$ multiplied by the
then-unchanged $P(x,E,t)$.  The second term is the probability density
$P(y,E,t)$ that the particle had some longitudinal momentum fraction $y>x$
at
time $t$, convolved with the probability that a splitting
took it from $y$ to $x$ during the small time interval $[t,t{+}\Delta t]$.
Doing the $y$ integral, and
re-arranging terms and taking the limit $\Delta t \to 0$, gives
\begin {equation}
   \frac{\partial}{\partial t} \, P(x,E,t)
   =
   - \Gamma(xE) \, P(x,E,t)
   +
   \int_{x}^1 \frac{d\xi}{\xi} \> \frac{d\Gamma}{d\xi}(\xi, \tfrac{x}{\xi}E)
   \, P(\tfrac{x}{\xi},E,t) .
\end {equation}
This can be rewritten in the infrared-safe form
\begin {equation}
   \frac{\partial}{\partial t} \, P(x,E,t)
   =
   -
   \int_0^1 d\xi \>
   \left[
     \frac{d\Gamma}{d\xi}(\xi, xE) \, P(x,E,t)
     -
     \frac{1}{\xi}
     \frac{d\Gamma}{d\xi}(\xi, \tfrac{x}{\xi}E)
     \, P(\tfrac{x}{\xi},E,t)
   \right] ,
\label {eq:Peq}
\end {equation}
where, in order to compactly combine integrals, we have adopted
the physically-appropriate convention that
\begin {equation}
   P(x,E,t) \equiv 0 ~~ \mbox{for $x>1$} .
\label {eq:Prange}
\end {equation}
The initial condition on the time-evolution (\ref{eq:Peq})
is that the particle start with energy $E$:
\begin {equation}
   P(x,E,0) = \delta(1-x) .
\end {equation}

As noted in the similar context of
ref.\ \cite{Nevolve1}, the distribution $P(x,E,t)$ will have
a $\delta$-function piece corresponding to stopped particles:
\begin {equation}
   P(x,E,t) = P_{\rm stopped}(E,t) \, \delta(x) + P_{\rm unstopped}(x,E,t) ,
\end {equation}
where $P_{\rm stopped}(E,t)$ in our case is the chance that the original
charged particle has already stopped by time $t$.
One can use this to recover the charge deposition distribution
$\rho(x,z)$ that will be needed to calculate, for example,
$\sigma/\lstop$.  The total probability that the charge is still moving
at time $t$ is
\begin {equation}
   Q_{\rm unstopped}(E,t) = \int_{0^+}^1 dx \> P(x,E,t) ,
\end {equation}
where $0^+$ represents a positive infinitesimal.  The charge deposition
distribution $\rho(E,z)$ [where $t$ and $z$ are interchangeable in our
discussion] is then
\begin {equation}
   \rho(E,t) = -\frac{\partial Q_{\rm unstopped}}{\partial t} \,.
\end {equation}
The moments $\langle z^n \rangle$
of that distribution are then (integrating once by parts)
\begin {equation}
  \langle z^n \rangle
  = \int_0^\infty dt \> t^n \rho(E,t)
  = n \int_0^\infty dt \> t^{n-1} \, Q_{\rm unstopped}(E,t)
  = n \int_0^\infty dt \> t^{n-1} \int_{0^+}^1 dx \> P(x,E,t) .
\label {eq:znfromP}
\end {equation}


\subsection {Leading-Order Equation}

As in the main text, assume that the {\it leading}-order splitting rate
scales with energy as $E^{-1/2}$ without any logs.  Even though we
then already know how to directly calculate the leading-order values of
the moments $\langle z^n \rangle$ using the method of
section \ref{sec:lstopNoLog}, we will still, it turns out, need to know the
leading-order solution
$P^{(0)}(x,E,t)$ to (\ref{eq:Peq}) in order to find the NLO correction
$P^{\rm(NLO)}(x,E,t)$ to $P$, which we will need to find the NLO correction
to moments.  So let's first discuss how to find that leading-order solution.
Similar to refs.\ \cite{Nevolve1,Nevolve2},
we can use the $E^{-1/2}$ scaling of the leading-order splitting rate
(and therefore $E^{1/2}$ scaling  of distance and time scales) to
simplify our basic equation (\ref{eq:Peq}) by writing
\begin {equation}
   P^{(0)}(x,E,t) = {\tilde P}^{(0)}(x, E^{-1/2} t)
   \equiv {\tilde P}^{(0)}(x, \tilde t) .
\label {eq:P0scale}
\end {equation}
Then (\ref{eq:Peq}) becomes%
\footnote{
  Our (\ref{eq:tildePeq})
  is similar in form to eq.\ (4) of ref.\ \cite{Nevolve1}.
  One difference is that their analysis
  accounts for all daughters of each splitting
  $g \to gg$ of a gluon shower because they are interested in the
  number distribution of gluons.  We would also need to do this to
  compute
  what we call ``energy'' deposition by this method.  But here we only need
  to track the fate of the single red-line particle in fig.\ \ref{fig:follow}
  to compute charge
  deposition $\rho(z,E)$ (at least in large-$N$ theories, as discussed
  in the main text).
  Additionally, there are some inessential
  normalization differences associated
  with how we define $d\tilde\Gamma$ from $d\Gamma$
  [related to their ${\cal K}(z)$ from $d{\cal P}_{\rm br}/dz\,d\tau$]
  and how we define our $\tilde t$ from $t$ [related
  to their $\tau$ from $t$].
}
\begin {equation}
   \frac{\partial}{\partial\tilde t} {\tilde P}^{(0)}(x,\tilde t)
   =
   -
   x^{-1/2}
   \int_{x}^1 d\xi \>
   \left[
     \frac{d\tilde\Gamma^{(0)}}{d\xi}(\xi) \, {\tilde P}^{(0)}(x,\tilde t)
     -
     \xi^{-1/2}
     \, {\tilde P}^{(0)}(\tfrac{x}{\xi},\tilde t)
   \right]
\label {eq:tildePeq}
\end {equation}
with initial condition ${\tilde P}^{(0)}(x,0) = \delta(1-x)$.
We have not found any closed form solutions, but
(\ref{eq:tildePeq}) could be solved numerically.
We will assume one has such a solution in hand and now discuss how
to obtain from it the NLO correction.


\subsection {Finding \boldmath$P^{\rm(NLO)}$ from \boldmath$P^{(0)}$}

Expand
\begin {equation}
  P(x,E,t) \simeq P^{(0)}(x,E,t) + P^{\rm(NLO)}(x,E,t) .
\label {eq:Pexpand}
\end {equation}
Making no assumptions about the energy dependence of the NLO piece
$d\Gamma^{\rm(NLO)}/d\xi$ of the splitting rate, we now show
how to determine
$P^{\rm(NLO)}$ from $P^{(0)}$.

Rather than returning to the general equation (\ref{eq:Peq}) for
$P(x,E,t)$, we find it easier to find $P^{\rm NLO}$ by a direct
probability argument.  First, as a matter of language, let's
call $d\Gamma^{(0)}/d\xi$ the rate for leading-order splittings
and $d\Gamma^{\rm(NLO)}/d\xi$ the rate for ``NLO splittings''
(e.g. the correction from overlapping double splittings, or the
virtual correction to single splittings).
Then imagine a shower composed of any number of leading-order splittings
plus zero or one or more NLO splittings, and correspondingly decompose
\begin {equation}
  P(x,E,t) \simeq
  P_{\rm(0~NLO~splits)}(x,E,t)
  + P_{\rm(1~NLO~split)}(x,E,t) .
\label {eq:Pdecompose}
\end {equation}
This is not quite the same as (\ref{eq:Pexpand}) because
$P^{(0)}$ is the probability if the {\it only} type of splitting possible
were leading-order splittings.
$P_{\rm(0~NLO~splits)}$, in contrast, imagines a world in which NLO
splittings are also possible but, by random chance, none have occurred
in the time interval $[0,t]$.  It differs from $P^{(0)}$ by having
to account for the probability that no NLO splittings occurred.

\begin {figure}[t]
\begin {center}
  \includegraphics[scale=0.75]{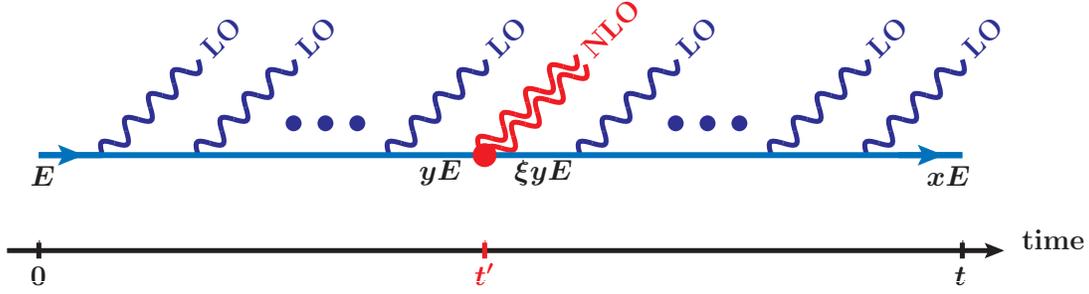}
  \caption{
     \label{fig:1NLO}
     A picture of following the original charge through a shower for a case
     where exactly one of the splittings involving that charge
     is a next-leading-order splitting.
  }
\end {center}
\end {figure}

Before examining the difference between $P^{(0)}$ and
$P_{\rm(0~NLO~splits)}$ in more detail, it will be helpful to first
figure out $P_{\rm(1~NLO~split)}$, depicted in fig.\ \ref{fig:1NLO}.
In formulas, this picture translates to
\begin {multline}
   P_{\rm(1~NLO~split)}(x,E,t)
   = \int_0^t dt' \int_x^1 dy \int_{x/y}^1 d\xi \>
        P_{\rm(0~NLO~splits)}(y, E, t') \,
        \frac{d\Gamma^{\rm(NLO)}}{d\xi}(\xi,yE) \,
\\ \times
        \left[
          (\xi y)^{-1} \, P_{\rm(0~NLO~splits})(\tfrac{x}{\xi y}, \xi y E, t-t')
        \right]
   .
\label {eq:P1NLO0}
\end {multline}
The first factor in the integrand gives the probability of making it
to time $t'$ with only leading-order splittings which take the energy
of the original particle down from $E$ to $yE$.  The second factor
gives the probability that there is then a NLO splitting in the
interval $[t',t'{+}dt']$ that takes the energy from $yE$ to $\xi y E$.
The final factor (in square brackets) is
related to the probability of then making it
from there to the later time $t$ with only leading-order splittings
taking the energy from $\xi y E$ to the final value
$x E = (x/\xi y)\times(\xi y E)$.  The reason for the normalization
factor $(\xi y)^{-1}$ inside the square brackets is because
the left-hand side of (\ref{eq:P1NLO0}) represent a probability
distribution for $x$ whereas the
$P_{\rm(0~NLO~splits)}(\tfrac{x}{\xi y}, \xi y E, t-t')$ on the right-hand
side represents a probability distribution for $x/\xi y$, and so
one must convert.

In (\ref{eq:P1NLO0}), the explicit factor $d\Gamma^{\rm(NLO)}/d\xi$ is
already next-to-leading order in $\alpha$.  Since we
are ultimately only interested in the expansion (\ref{eq:Pexpand})
of probabilities characterizing the shower to
next-to-leading-order, we may therefore
approximate the other factors $P_{\rm(0~NLO~splits)}$ in
(\ref{eq:P1NLO0}) by $P^{(0)}$.
So
\begin {multline}
   P_{\rm(1~NLO~split)}(x,E,t)
   \simeq \int_0^t dt' \int_x^1 dy \int_{x/y}^1 d\xi \>
        P^{(0)}(y, E, t') \,
        \frac{d\Gamma^{\rm(NLO)}}{d\xi}(\xi,yE) \,
\\ \times
        \left[
          (\xi y)^{-1} \, P^{(0)}(\tfrac{x}{\xi y}, \xi y E, t-t')
        \right]
   .
\label {eq:P1NLO}
\end {multline}

\begin {figure}[t]
\begin {center}
  \includegraphics[scale=0.75]{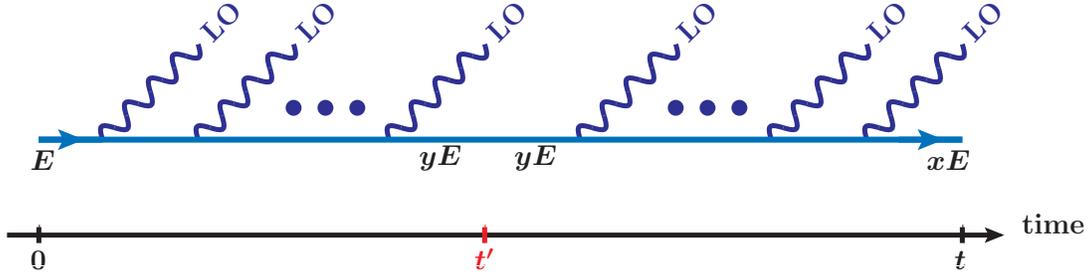}
  \caption{
     \label{fig:0NLO}
     Like fig.\ \ref{fig:1NLO} but showing a case where (randomly) all of
     the splittings are leading-order splittings.  To evaluate its
     probability, we must combine the probabilities that in each tiny
     time interval $[t',t'{+}dt']$ there was no NLO splitting.
  }
\end {center}
\end {figure}

Now let's return to the first term of (\ref{eq:Pdecompose}),
$P_{\rm(0~NLO~splits)}$, and its difference from $P^{(0)}$ at first
order in $d\Gamma^{\rm(NLO)}$.  To evaluate this difference,
consider fig.\ \ref{fig:0NLO}, which is the analog of
fig.\ \ref{fig:1NLO} but with {\it no} NLO splitting occurring
during the time interval $[t',t'{+}dt']$.
The corresponding formula analogous to (\ref{eq:P1NLO}) is
\begin {multline}
   P_{\rm(0~NLO~splits)}(x,E,t)
   \simeq P^{(0)}(x,E,t)
     - \int_0^t dt' \int_x^1 dy \int_0^1 d\xi \>
        P^{(0)}(y, E, t') \,
        \frac{d\Gamma^{\rm(NLO)}}{d\xi}(\xi,yE) \,
\\ \times
        \left[
          y^{-1} \, P^{(0)}(\tfrac{x}{y}, y E, t-t')
        \right]
   ,
\label {eq:P0NLO}
\end {multline}
where
\begin {equation}
  dt' \> \int_0^1 d\xi \>
        \frac{d\Gamma^{\rm(NLO)}}{d\xi}(\xi,yE) \,
\end {equation}
represents the probability of any NLO splitting that could have happened
in the time interval $[t',t'+dt']$ but didn't.

Combining (\ref{eq:P1NLO}) and (\ref{eq:P0NLO}) to compare
(\ref{eq:Pdecompose}) with (\ref{eq:Pexpand}), we then have
\begin {multline}
   P^{\rm(NLO)}(x,E,t)
   =
   - \int_0^t dt' \int_x^1 dy \int_0^1 d\xi \>
        P^{(0)}(y, E, t') \,
        \frac{d\Gamma^{\rm(NLO)}}{d\xi}(\xi,yE) \,
\\ \times
        \left[
          y^{-1} \, P^{(0)}(\tfrac{x}{y}, y E, t-t')
          -
          (\xi y)^{-1} \, P^{(0)}(\tfrac{x}{\xi y}, \xi y E, t-t')
        \right]
   ,
\label {eq:PNLO}
\end {multline}
where, in order to combine integrals, we have again adopted the
convention (\ref{eq:Prange}).


\subsection {NLO corrections to moments \boldmath$\langle z^n\rangle$}

Similar to our notation $\lstop \simeq \lstop^{(0)} + \Delta \lstop$,
we'll write the expansion of other moments of the charge deposition
distribution $\rho(z,E)$ to
NLO as
\begin {equation}
  \langle z^n \rangle \simeq
  \langle z^n \rangle^{(0)} + \Delta\langle z^n \rangle .
\end {equation}
Putting (\ref{eq:PNLO}) into (\ref{eq:znfromP}),
\begin {multline}
  \Delta\langle z^n \rangle =
  -
  n \int_0^\infty dt \> t^{n-1} \int_{0^+}^1 dx
   \int_0^t dt' \int_x^1 dy \int_0^1 d\xi \>
        P^{(0)}(y, E, t') \,
        \frac{d\Gamma^{\rm(NLO)}}{d\xi}(\xi,yE) \,
\\ \times
        \left[
          y^{-1} \, P^{(0)}(\tfrac{x}{y}, y E, t-t')
          -
          (\xi y)^{-1} \, P^{(0)}(\tfrac{x}{\xi y}, \xi y E, t-t')
        \right]
   .
\label {eq:PznA}
\end {multline}
Because $d\Gamma^{(0)}$ and $P^{(0)}$ have simple scaling with $E$, we will
find it convenient to scale out the same factors of $E^{1/2}$ from the
NLO rate as in the main text:
\begin {equation}
   \frac{d\tilde\Gamma^{\rm(NLO)}}{d\xi}(\xi,E) \equiv
   E^{-1/2} \, \frac{d\tilde\Gamma^{\rm(NLO)}}{d\xi}(\xi,E) .
\end {equation}
Using (\ref{eq:P0scale}), eq.\ (\ref{eq:PznA}) above can be written
\begin {multline}
  \Delta\langle \tilde z^n \rangle =
  -
  n \int_0^\infty d\tilde t \> \tilde t^{n-1} \int_{0^+}^1 dx
   \int_0^{\tilde t} d\tilde t' \int_x^1 dy \int_0^1 d\xi \>
        \tilde P^{(0)}(y, \tilde t') \,
        y^{-1/2} \,
        \frac{d\tilde\Gamma^{\rm(NLO)}}{d\xi}(\xi,yE) \,
\\ \times
        \left[
          y^{-1} \,
          \tilde P^{(0)}\bigl(\tfrac{x}{y}, y^{-1/2}(\tilde t-\tilde t')\bigr)
          -
          (\xi y)^{-1} \,
          \tilde P^{(0)}\bigl(\tfrac{x}{\xi y},
              (\xi y)^{-1/2}(\tilde t-\tilde t')\bigr)
        \right]
   .
\label {eq:PznAtilde}
\end {multline}

Changing integration variable from $\tilde t$ to
$\tilde T \equiv \tilde t-\tilde t'$,
we may rewrite
\begin {multline}
   \int_0^\infty d\tilde t \> {\tilde t}^{n-1}
      \int_0^{\tilde t} d{\tilde t}' \cdots
   =
   \int_0^\infty d\tilde T \int_0^\infty d{\tilde t}'
      \> ({\tilde t}'+\tilde T)^{n-1} \cdots
\\
   =
   \sum_{k=0}^{n-1}
   \begin{pmatrix} n{-}1 \\ k \end{pmatrix}
   \int_0^\infty d\tilde T \> {\tilde T}^k \>
      \int_0^\infty d{\tilde t}' \> ({\tilde t}')^{n-1-k} \cdots
\end {multline}
in (\ref{eq:PznAtilde}).
Defining the time-moments
\begin {equation}
   {\tilde P}^{(0)}_j(x)
   \equiv \int_0^\infty d\tilde t \> \tilde t^j \, {\tilde P}^{(0)}(x,\tilde t)
\label {eq:P0moment}
\end {equation}
of ${\tilde P}^{(0)}(x,t)$, we can then write
\begin {multline}
  \Delta\langle \tilde z^n \rangle =
  -n
  \sum_{k=0}^{n-1}
  \begin{pmatrix} n{-}1 \\ k \end{pmatrix}
  \int_{0^+}^1 dx \int_x^1 dy \int_0^1 d\xi \>
        \tilde P^{(0)}_{n-1-k}(y) \,
        \frac{d\tilde\Gamma^{\rm(NLO)}}{d\xi}(\xi,yE) \,
\\ \times
        y^{-1/2}
        \left[
          y^{(k-1)/2} \tilde P^{(0)}_k\bigl(\tfrac{x}{y}\bigr)
          -
          (\xi y)^{(k-1)/2} \,
          \tilde P^{(0)}_k\bigl(\tfrac{x}{\xi y}\bigr)
        \right]
   .
\label {eq:PznB}
\end {multline}
The constraint (\ref{eq:Prange}) allows us to
replace the lower limit of integration on the $y$ integral by zero.
Continuing to use (\ref{eq:Prange}),
the $x$ integration  in (\ref{eq:PznB}) can then be accomplished
as follows:
\begin {equation}
   \int_{0^+}^1 dx \> \tilde P^{(0)}_k\bigl( \tfrac{x}{\xi y} \bigr)
   = \xi y \int_{0^+}^1 d\bar x \> P^{(0)}_k(\bar x)
   = \frac{\xi y}{k+1} \langle \tilde z^{k+1} \rangle^{(0)} ,
\end {equation}
where the last equality follows from (\ref{eq:znfromP})
and (\ref{eq:P0moment}).
So, with a little manipulation and switching summation
variable to $p\equiv k{+}1$,
(\ref{eq:PznB}) becomes
\begin {equation}
  \Delta\langle \tilde z^n \rangle =
  -
  \sum_{p=1}^{n}
  \begin{pmatrix} n \\ p \end{pmatrix}
  \langle \tilde z^p \rangle^{(0)}
  \int_0^1 dy \> y^{(p-1)/2} \, \tilde P^{(0)}_{n-p}(y) \,
  \int_0^1 d\xi \> \frac{d\tilde\Gamma^{\rm(NLO)}}{d\xi}(\xi,yE) \,
       [1-\xi^{p/2}]
  ,
\label {eq:PznC}
\end {equation}
where the leading-order
$\langle \tilde z^n \rangle^{(0)}$ is given recursively by the
simple result of section \ref{sec:lstopNoLog} applied to
the leading-order rate:
\begin {equation}
   \langle \tilde z^p \rangle^{(0)} =
   \frac{ p \langle \tilde z^{p-1} \rangle^{(0)} }
        { \int_0^1 d\xi \> \frac{d\tilde\Gamma^{(0)}}{d\xi} [1-\xi^{p/2}] }
   \,.
\end {equation}
(\ref{eq:PznC}) is our result for how to isolate the NLO correction to
the moments of $\rho(E,z)$ from a numerical calculation of
the leading-order $\tilde P^{(0)}(z,t)$, for any energy dependence of
$d\Gamma^{\rm(NLO)}/d\xi$.
We leave the matter there.%
\footnote{
  We have checked that in the special case where the {\it only} energy
  dependence of $d\Gamma^{\rm(NLO)}/d\xi$ is an overall factor of
  $E^{-1/2}$
  (i.e.\ where $d\tilde\Gamma^{\rm(NLO)}/d\xi$ is independent of energy),
  then the factorization of the integrals in (\ref{eq:PznC}) allows one
  to show, after a somewhat lengthy argument,
  that this formula reproduces the answer one would get
  by expanding the result of section \ref{sec:lstopNoLog}
  for $\langle z^n \rangle$.
}


\section{Limiting behavior and numerical interpolation of
         \boldmath$d\Gamma/d\xe\,d\yfrake$}
\label {app:interpolate}

In this appendix, we discuss limiting cases of the contributions
of diagrams (f) and (k) to $d\Gamma/d\xi\,d\yfrake$, and then use that
knowledge to smooth out those contributions into forms that can
be more easily interpolated numerically.


\subsection{\boldmath$2\Re[d\Gamma^{\rm(subtracted)}/d\xe\,d\yfrake]$}

\begin {figure}[t]
\begin {center}
  \begin{picture}(450,155)(0,0)
  \put(0,15){\includegraphics[scale=0.55]{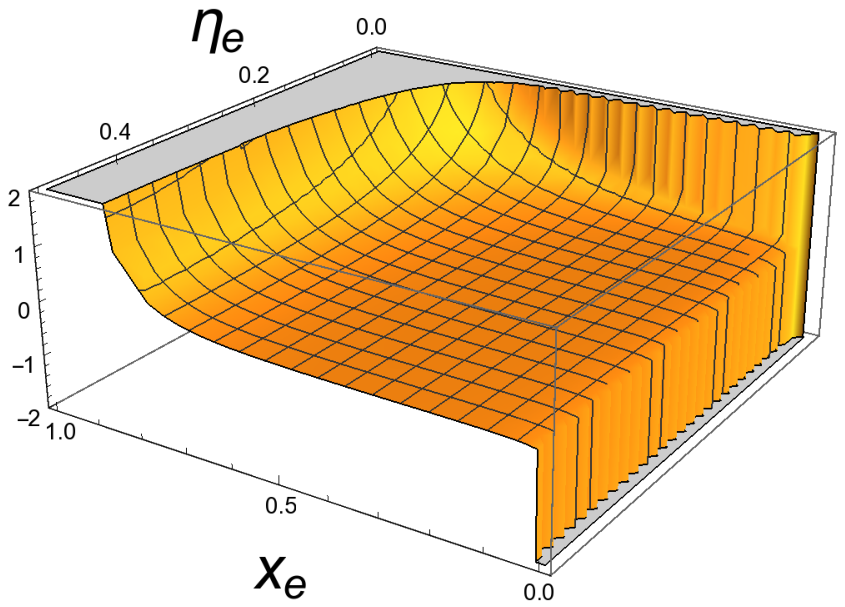}}
  \put(153,15){\includegraphics[scale=0.55]{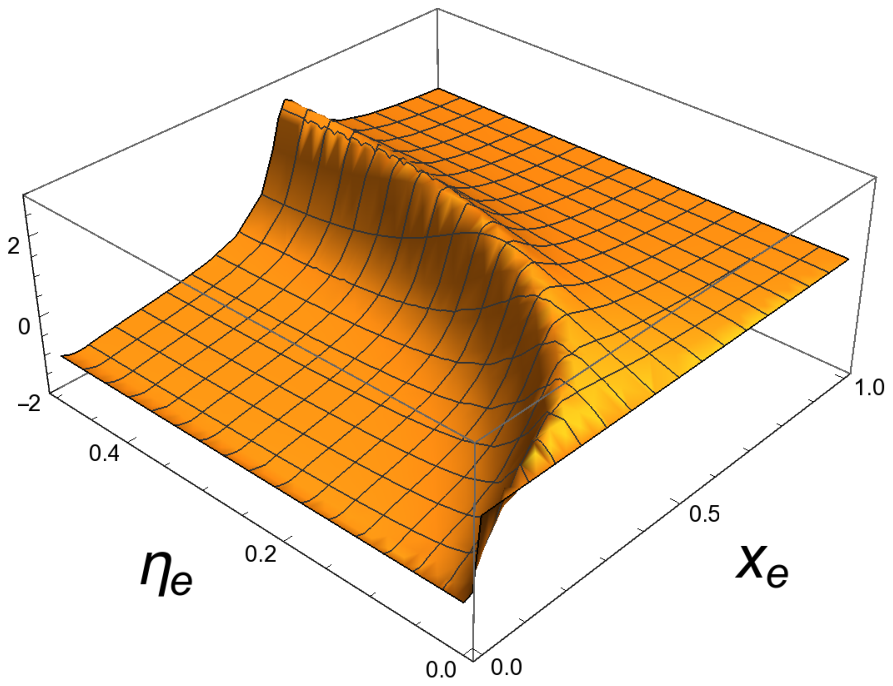}}
  \put(306,2){\includegraphics[scale=0.55]{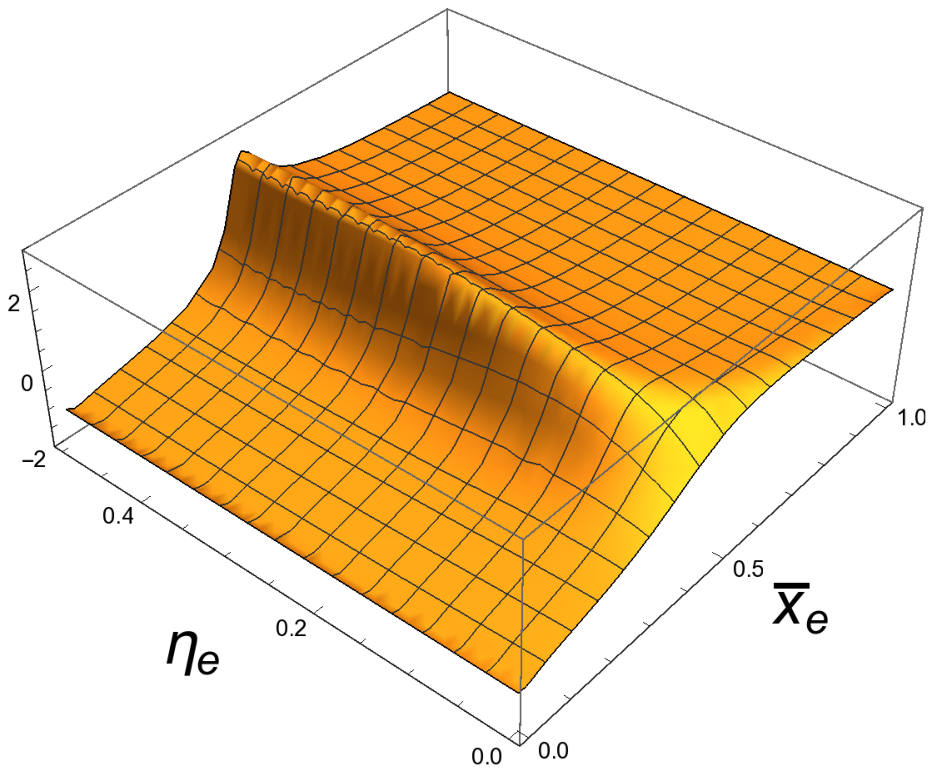}}
  \put(70,0){\scriptsize(a)}
  \put(223,0){\scriptsize(b)}
  \put(376,0){\scriptsize(c)}
  \end{picture}
  \caption{
     \label{fig:interpolate}
     (a) The original function $f(\xe,\yfrake)$ of (\ref{eq:fdef}),
     which diverges in various limits;
     (b) the corresponding finite function $g$ of (\ref{eq:gdef});
     (c) a smoothed version of $g$ created by transforming $\xe$
     to ${\bar x}_e$ as in (\ref{eq:barxe}).
  }
\end {center}
\end {figure}

For simplicity of notation, define
\begin {equation}
   f(\xe,\yfrake) \equiv
   2\Re\left[\frac{d\Gamma}{d\xe\,d\yfrake}\right]^{\rm(subtracted)}_{xyy\bar x}
   =
   (1{-}\xe) \,
   2\Re\left[\frac{d\Gamma}{d\xe\,d\ye}\right]^{\rm(subtracted)}_{xyy\bar x}
\label {eq:fdef}
\end {equation}
Fig.\ \ref{fig:interpolate}a show a plot of $f(\xe,\yfrake)$.
Both $\xe$ and $\yfrake$ run from 0 to 1, but in large-$\Nf$ QED
the result for $d\Gamma/d\xi$ is symmetric under interchanging the
two daughters of the photon pair production, corresponding to
symmetry under $\yfrake \to 1{-}\yfrake$.  Taking advantage of this,
we will only plot, discuss, and interpolate $f(\xe,\yfrake)$
here and in what follows for
\begin {equation}
   0 < \xe < 1, \qquad 0 < \yfrake \le \tfrac12 .
\end {equation}

The function in fig.\ \ref{fig:interpolate}a diverges to $+\infty$
for $x{\to}1$ and $\eta{\to}0$ and diverges to $-\infty$ for
$x{\to 0}$.  Section II of Ref.\ \cite{QEDnf} found that
the $x{\to}1$ divergence of the {\it double} splitting process
$e \to e\bar e e$ is
\begin {equation}
   \left[ \frac{d\Gamma}{d\xe\, d\yfrake} \right]_{e\to e\bar e e}
   \approx
   \frac{\alphaqed}{2\pi} \, P_{e\to e}(\xe) \,
   \ln\Bigl( \frac{t_{{\rm form},\xe}}{t_{{\rm form},\yfrake}} \Bigr) \,
   \left[ \frac{d\Gamma}{d\yfrake} \right]_{\gamma \to e\bar e}^{\rm LO} ,
\label {eq:eeeeLimit}
\end {equation}
at leading-log order,
where (also at leading-log order) one may use parametric estimates for the
formation times in the argument of the logarithm.
We find numerically that this same formula works for
the NLO single-splitting correction (\ref{eq:fdef}).
Moreover, we find it works whenever
the formation time $t_{{\rm form},\yfrake}$ for pair production from the
photon is parametrically small compared to the formation time
$t_{{\rm form},\xe}$ for the bremsstrahlung, which includes the
limit $\yfrake \to 0$ as well as $\xe \to 1$.
Using numerics, we have also extracted the constant under the logarithm
for this limit, finding that
$f(\xe,\yfrake)$
approaches
\begin {equation}
  F_1(\xe,\yfrake) \equiv
  \frac{\alphaqed}{2\pi} \, P_{e\to e}(\xe) \,
  \left[
     \ln\Bigl( \frac{\xe^{1/2}}{(1-\xe)[\yfrake(1{-}\yfrake)]^{1/2}} \Bigr)
     - 1.2607
  \right]
  \left[ \frac{d\Gamma}{d\yfrake} \right]_{\gamma \to e\bar e}^{\rm LO}
\end {equation}
both for (i) $\xe{\to}1$ and for (ii) $\yfrake{\to}0$ with $\xe$ fixed.
Above,
\begin {equation}
   \left[ \frac{d\Gamma}{d\yfrake} \right]_{\gamma \to e\bar e}^{\rm LO}
   =
   \frac{\Nf\alphaqed}{\pi} P_{\gamma\to e}(\yfrake)
      \, \Re(i \Omega_0^{\gamma\to e\bar e}) ,
\end {equation}
\begin {equation}
   \Omega_0^{\gamma\to e\bar e}
    = \sqrt{ \frac{-i \hat q}{2\yfrake(1{-}\yfrake) E_\gamma} }
    = \sqrt{ \frac{-i \hat q}{2\yfrake(1{-}\yfrake) x_\gamma E} } \,,
\label {eq:Omega0pair}
\end {equation}
\begin {equation}
  P_{e\to e}(\xe) = \frac{1 + \xe^2}{1-\xe} ,
  \qquad
  P_{\gamma\to e}(\yfrake) = \yfrake^2 + (1-\yfrake)^2 .
\end {equation}

For the $\xe{\to}0$ divergence, we do not know
of a physical argument like the one of ref.\ \cite{QEDnf} for the
previous limit (\ref{eq:eeeeLimit}), and we have not tried to derive
this limit analytically from the general formula for $f(\xe,\yfrake)$.
However, we have found by
numerical experimentation that $f(\xe,\yfrake)$ approaches
\begin {equation}
  F_0(\xe,\yfrake) \equiv
  -
  {\Nf\alphaqed^2}{c_0} \, P_{\gamma\to e}(\yfrake) \,
  \sqrt{\frac{1{-}\xe}{\xe}}
\end {equation}
for $\xe{\to}0$ with $\yfrake$ fixed,
where $c_0$ is a constant whose numerical value is approximately
\begin {equation}
  c_0 \cong \frac{1}{10 \pi} \,.
\end {equation}

Next, we combined the limits $F_0$ and $F_1$ to make a positive
weighting function $F$ that correctly approximates the magnitude
of $f$ in all of the divergent limiting cases.  The goal was to
create a new function
\begin {equation}
   g(\xe,\yfrake) \equiv \frac{f(\xe,\yfrake)}{F(\xe,\yfrake)}
\label {eq:gdef}
\end {equation}
that would not be divergent.  By trial and
error, trying to choose a $g$ that looked reasonable, we settled
on the choice
\begin {equation}
   F(\xe,\yfrake) \equiv
   \sqrt{ [F_1(\xe,\yfrake)]^2 - F_1(\xe,\yfrake) \, F_0(\xe,\yfrake)
        + [F_0(\xe,\yfrake)]^2 } .
\end {equation}
The resulting finite function $g(\xe,\yfrake)$ is shown in fig.
\ref{fig:interpolate}b.  Note that it has an unfortunate
finite directional singularity for $(\xe,\yfrake) \to (0,0)$,
where the value of $g$ depends on how one approaches the limit.
One can trace this back to a similar directional singularity in
how $f$ diverged as $(\xe,\yfrake) \to (0,0)$ in fig.\ \ref{fig:interpolate}a.

The directional singularity in fig.\ \ref{fig:interpolate}b complicates
good numerical interpolation of $g$ from a finite mesh
of values.  We therefore looked for a $\yfrake$-dependent
change of the $\xe$ variable, $\xe \to {\bar x}_e(\yfrake)$, that would
map the interval $[0,1]$ into itself while shifting the location of
the ridge in fig.\ \ref{fig:interpolate}b to avoid running into the
corner.  To map $[0,1]$ into itself, we considered a transformation
of the form
\begin {subequations}
\label{eq:barxe}
\begin {equation}
  {\bar x}_e(\yfrake) =
  \frac{\bigl(1-x_0(\yfrake)\bigr) \xe}
       {x_0(\yfrake)+\bigl(1-2x_0(\yfrake)\bigr)\xe} \,,
\label {eq:xbardef}
\end {equation}
which maps $\xe=x_0(\yfrake)$ to ${\bar x}_e=\frac12$.  By experimentation,
we chose
\begin {equation}
   x_0(\yfrake) =
   \left(a + \frac{b}{\yfrake}\right) \frac{\yfrake^2}{(\yfrake+c)^2}
\end {equation}
with
\begin {equation}
   a = 0.63775522, \qquad b = 0.1264279, \qquad c=0.09664453 .
\end {equation}
\end {subequations}
Fig.\ \ref{fig:interpolate}c then shows $g$ as a function of
$({\bar x}_e,\yfrake)$ instead of $(\xe,\yfrake)$.  This function
is smooth enough for numerical interpolation.  Our procedure was
to evaluate $f$ for a mesh of points in $({\bar x}_e,\yfrake)$ space,
from which we obtain $g$ on that mesh of points, and then interpolate to get
$g$ in the entire $({\bar x}_e,\yfrake)$ region.  The original function
$f(\xe,\yfrake)$ can then be evaluated from this interpolation by
inverting all of the steps described above.


\subsection{\boldmath$2\Re[2\, d\Gamma_{(f)}/d\xe\,d\yfrake]$}

\begin {figure}[t]
\begin {center}
  \begin{picture}(450,155)(0,0)
  \put(0,15){\includegraphics[scale=0.55]{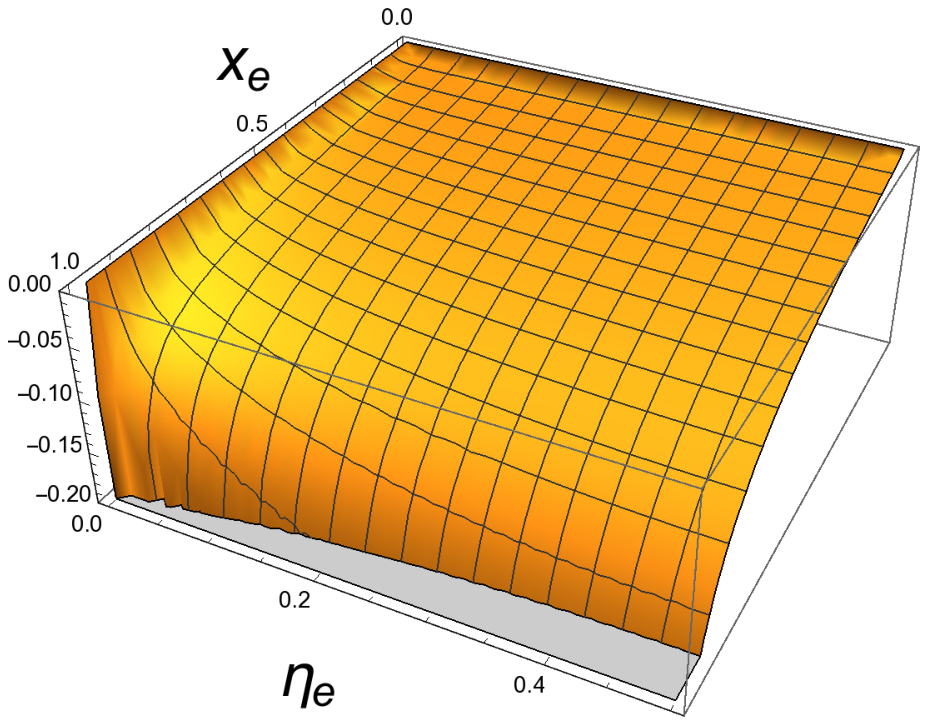}}
  \put(153,15){\includegraphics[scale=0.55]{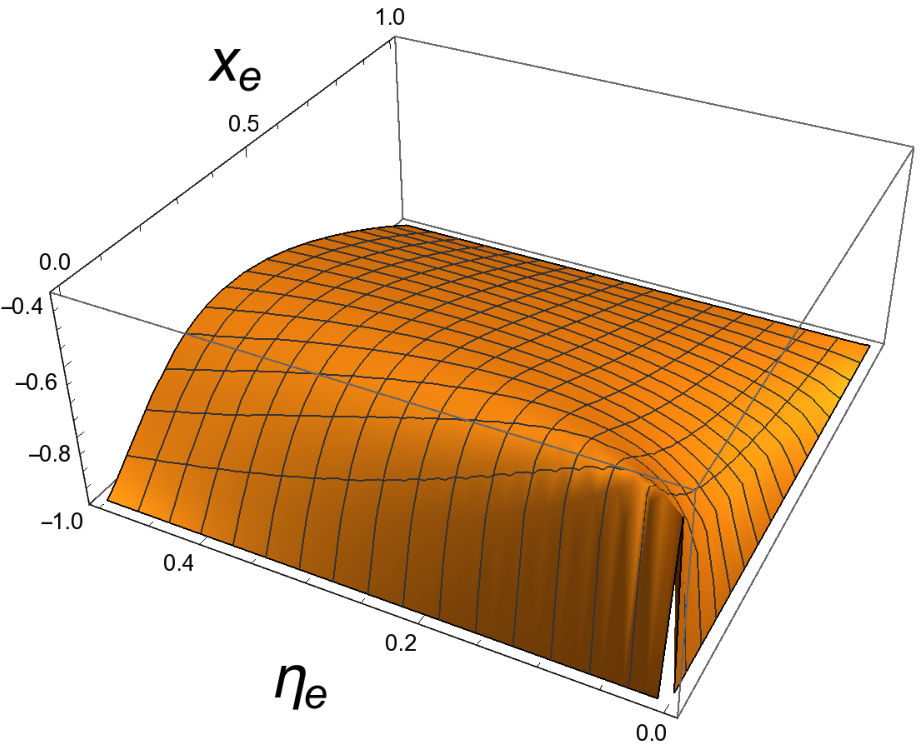}}
  \put(306,15){\includegraphics[scale=0.55]{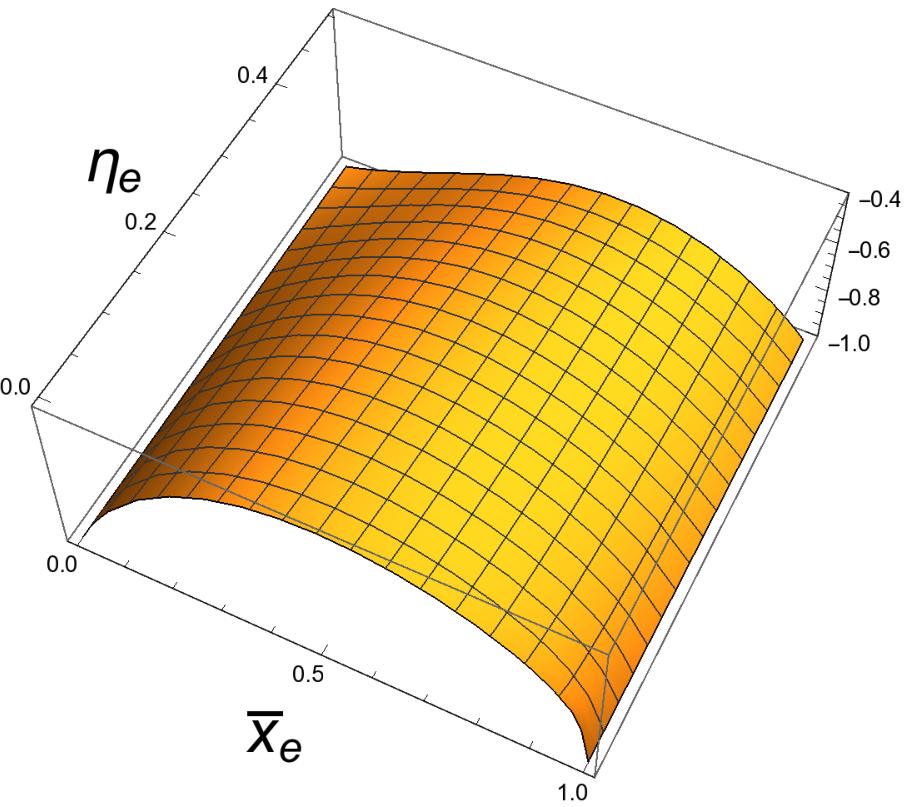}}
  \put(70,0){\scriptsize(a)}
  \put(223,0){\scriptsize(b)}
  \put(376,0){\scriptsize(c)}
  \end{picture}
  \caption{
     \label{fig:interpolateI}
     Like fig. \ref{fig:interpolate} but for the $f(\xe,\yfrake)$
     of (\ref{eq:fdefI}).
  }
\end {center}
\end {figure}

Fig.\ \ref{fig:interpolateI} shows the same sort of transformations
but here for the contribution
\begin {equation}
   f(\xe,\yfrake) \equiv
   2\times 2\Re\left[\frac{d\Gamma}{d\xe\,d\yfrake}\right]_{\rm(f)}
   =
   2\times (1{-}\xe) \, 2\Re\left[\frac{d\Gamma}{d\xe\,d\ye}\right]_{\rm(f)}
\label {eq:fdefI}
\end {equation}
to $d\Gamma/d\xi$ in (\ref{eq:GammaNLOsum2}).

One difference here is that the only divergence of $f$ is for $\xe{\to}1$.
However, there is $\sqrt{\yfrake}$ or $\sqrt{\xe}$ behavior in other limits,
which is best to also remove before interpolation.  So we again look
at all the limiting cases.  We find
\begin {equation}
  F_1(\xe,\yfrake) \equiv
  -  \frac{\Nf\alphaqed^2}{\pi^2} \, (1{+}\xe)
  \left(
     \frac{\yfrake(1-\yfrake) \hat q}{(1-\xe) E}
  \right)^{1/2}
  \ln 2 ,
\end {equation}
\begin {equation}
  F_0(\xe,\yfrake) \equiv
  -
  \frac{\Nf\alphaqed^2}{\pi^2}
  \left(
     \frac{\xe \hat q}{E}
  \right)^{1/2}
  c'_0 ,
\end {equation}
with
\begin {equation}
  c'_0 =
   \int_0^\infty du \>
   \frac{u-\operatorname{th}u}
        {u^2 (1+\operatorname{th}u)\operatorname{sh}u}
  =
  0.306853 \,,
\end {equation}
and we choose
\begin {equation}
   F(\xe,\yfrake) \equiv
   \sqrt{
     \frac{\xe^2[F_1(\xe,\yfrake)]^2 + \yfrake^2(1{-}\xe)^2[F_0(\xe,\yfrake)]^2}
          {\xe^2 + \yfrake^2(1{-}\xe)^2}
   }
\end {equation}
to get fig.\ \ref{fig:interpolateI}b.
Finally, here we take (\ref{eq:xbardef}) with
\begin {equation}
   x_0(\eta) =
   1.2 \, \yfrake (1{-}\yfrake)
\end {equation}
to get fig.\ \ref{fig:interpolateI}c.

Any reader going through this appendix may be forgiven if they think it's
all a bit convoluted.  We agree and would be happy to have a more efficient
procedure (in terms of our time or the computer's).


\end {document}